%% file: main.tex
\documentclass{IEEEtran}
\usepackage{cite}
\usepackage{amsmath,amssymb,amsfonts}
\usepackage{algorithmic}
\usepackage{graphicx,color}
\usepackage{textcomp}

\usepackage{anyfontsize}
\usepackage{tabulary}
\usepackage[table]{xcolor}
\usepackage{multirow,multicol}
\usepackage{array,ragged2e}
\usepackage{tikz,soul}
\usepackage[bookmarks=false]{hyperref}

\input{my-macro}

\def\BibTeX{{\rm B\kern-.05em{\sc i\kern-.025em b}\kern-.08em
    T\kern-.1667em\lower.7ex\hbox{E}\kern-.125emX}}
\AtBeginDocument{\definecolor{ojcolor}{cmyk}{0.93,0.59,0.15,0.02}}

\begin{document}

This work has been submitted to the IEEE for possible publication. Copyright may be transferred without notice, after which this version may no longer be accessible.

\newpage

\title{On the Coexistence of eMBB and URLLC in Multi-cell Massive MIMO}

\author{Giovanni Interdonato, {\em Member}, {\em IEEE}, Stefano Buzzi, {\em Senior Member}, {\em IEEE},  Carmen D'Andrea, {\em Member}, {\em IEEE}, 
	Luca Venturino, {\em Senior Member}, {\em IEEE}, Ciro D'Elia, and Paolo Vendittelli
\thanks{This work was supported by the Ministero delle Imprese e del Made in Italy (former MISE) within the project ``Smart Urban Mobility Management'' (5G-SUMMA), Asse II, Supporto alle Tecnologie Emergenti.}
\thanks{G. Interdonato, S. Buzzi, C. D'Andrea, L. Venturino and C. D'Elia are with the
	Department of Electrical and Information Engineering, University of Cassino and Southern Latium, 03043 Cassino, Italy. They are also affiliated with Consorzio Nazionale Interuniversitario per le Telecomunicazioni (CNIT), 43124 Parma, Italy. P. Vendittelli is with TIM S.p.A., 20133 Milan, Italy. S. Buzzi is also affiliated with 
Politecnico di Milano, 20133 Milan, Italy.}
\thanks{Corresponding author: Giovanni Interdonato.}}

\maketitle
\thispagestyle{empty}

\markboth{ON THE COEXISTENCE OF EMBB AND URLLC IN MULTI-CELL MASSIVE MIMO}{INTERDONATO \textit{et al.}} 

\begin{abstract}
The non-orthogonal coexistence between the enhanced mobile broadband (eMBB) and the ultra-reliable low-latency communication (URLLC) in the downlink of a multi-cell massive MIMO system is rigorously analyzed in this work. 
We provide a unified information-theoretic framework blending an infinite-blocklength analysis of the eMBB spectral efficiency (SE) in the ergodic regime with a finite-blocklength analysis of the URLLC error probability relying on the use of mismatched decoding, and of the so-called saddlepoint approximation. Puncturing (PUNC) and superposition coding (SPC) are considered as alternative downlink coexistence strategies to deal with the inter-service interference, under the assumption of only statistical channel state information (CSI) knowledge at the users. 
eMBB and URLLC performances are then evaluated over different precoding techniques and power control schemes, by accounting for imperfect CSI knowledge at the base stations, pilot-based estimation overhead, pilot contamination, spatially correlated channels, the structure of the radio frame, and the characteristics of the URLLC activation pattern.
Simulation results reveal that SPC is, in many operating regimes, superior to PUNC in providing higher SE for the eMBB yet achieving the target reliability for the URLLC with high probability. Moreover, PUNC might cause eMBB service outage in presence of high URLLC traffic loads. However, PUNC turns to be necessary to preserve the URLLC performance in scenarios where the multi-user interference cannot be satisfactorily alleviated.
\vspace*{3mm}
\end{abstract}



\newcounter{counter_1}

\maketitle

\section{INTRODUCTION}
\IEEEPARstart{W}{ith} the advent of the mobile application ecosystem and the resulting increase of the data-processing and storage capabilities of the smart devices, several heterogeneous services have emerged setting various stringent communication requirements in terms of data rates, latency, reliability and massive connectivity. These requirements and related use cases have been summarized by the 3rd Generation Partnership Project (3GPP) into three macro services, namely \textit{enhanced mobile broadband} (eMBB), ultra-reliable low-latency communications (URLLC) and \textit{massive machine-type communications} (mMTC)~\cite{ituvision2015}.
eMBB services require high-peak data-rate and stable connectivity, and include most of the everyday usage applications: entertainment, multimedia, communication, collaboration, mapping, web-surfing, etc.
URLLC services require an one-way radio latency of 1 ms with 99.999\% success probability, and include real-time and time-critical applications, such as autonomous driving, automation control, augmented reality, video and image processing, etc. mMTC services enable connectivity between a vast number of miscellaneous devices, and include applications such as smart grids, traffic management systems, environmental monitoring, etc. \vfill 

5G started to roll out variously as an eMBB service, essentially like a faster version of LTE, whereas mMTC and URLLC requirements continue to be refined and will materialize within the next decade, although some experimental activities are already taking place in many parts of the world\footnote{See, e.g., the funding programs from the Italian former Ministry of Economic Development, as well as those of other European Countries, the EU, USA, China and Japan.}. Academic research and industrial standardization is currently interested at different coexistence mechanisms for such heterogeneous services, apparently moving apart from the initial vision of a \textit{sliced} network~\cite{Popovski2018slicing}. 
Slicing the network basically means allocating orthogonal resources (storage, computing, radio communications, etc.) to heterogeneous services so that to guarantee their mutual isolation. This approach is, in broad sense, generally known as \textit{orthogonal multiple access} (OMA). As an interesting alternative to orthogonal resource allocation, non-orthogonal OMA (NOMA) is gaining increasing importance especially with respect to the allocation of the \textit{radio access network} (RAN) communication resources.      
The conventional approach to slice the RAN is to separate eMBB, mMTC, and URLLC services in time and/or frequency domains, whereas NOMA relies on efficient coexistence strategies wherein heterogeneous services share the same time-frequency resources, being separated in the power and spatial domain. In this regard, the terminology \textit{Heterogeneous} OMA (H-OMA) is often adopted~\cite{Popovski2018slicing} to distinguish the orthogonal resource allocation of heterogeneous services from that of the same type, referred to as OMA. (The same distinction applies to H-NOMA with respect to NOMA.) \vfill 

Massive MIMO~\cite{Marzetta2010,redbook,massivemimobook} is a technology that uses a very large number of co-located antennas at the base stations (BSs) to coherently and simultaneously serve multiple users over the same radio resources. The users are multiplexed in the spatial domain by using beamforming techniques that enable high-directivity transmission and reception. The use of many antennas also triggers the \textit{favorable propagation} which further reduces the multi-user interference and the \textit{channel hardening} which reduces the random fluctuations of the effective channel gain. As a consequence, there is no need to adopt intricate signal processing techniques to deal with the multi-user interference. Such an aggressive spatial multiplexing along with the intrinsic practicality and scalability of the massive MIMO technology leads to high levels of energy and spectral efficiency,  spatial diversity, link reliability and connectivity. 

The primary focus of the massive MIMO research has been on increasing the user data rates, thereby targeting the eMBB requirements. Lately, some studies have highlighted the significant benefits that massive MIMO is able to provide to URLLC~\cite{Popovski2018,Bana2019,Ostman2021} by reducing the outage and error probability, and therefore increasing the link reliability. Higher reliability results to less retransmissions which, in turn, translates to a lower latency. 
mMTC also benefits from massive MIMO technology~\cite{Bana2019, Bjornson2017} by capitalizing on the high energy efficiency to increase devices' battery lifetime. Besides, favorable propagation enables an aggressive spatial multiplexing of the mMTC devices, facilitating the detection and the random access procedures. 

\subsection{RELATED WORKS}
Coexistence between heterogeneous services has been initially studied in systems wherein a single-antenna BS serves multiple heterogeneous users. In~\cite{Popovski2018slicing}, Popovski \textit{et al.} proposed a first tractable communication-theoretic model that captures the key features of eMBB, URLLC and mMTC traffic. (These features are summarized in~\Tableref{tab:features}.) 
Specifically,~\cite{Popovski2018slicing} analyzes two scenarios for a \textit{single-cell} model: $(i)$ slicing for URLLC and eMBB, and $(ii)$ slicing for mMTC and eMBB. The downlink multiplexing of URLLC and eMBB is studied in~\cite{Anand2018} with the goal of maximizing the utility of the eMBB traffic while satisfying the quality of service requirements of the URLLC traffic, and by abstracting the operation at the physical layer. Coexistence mechanisms between URLLC and eMBB traffic, based on the \textit{puncturing} technique, have been proposed in~\cite{Kassab2019} for the uplink of a \textit{multi-cell} network wherein a simplified Wyner channel model with no fading was assumed.
As for multi-user MIMO systems, in~\cite{Esswie2018} a null-space-based spatial preemptive scheduler for joint URLLC and eMBB traffic is proposed for cross-objective optimization, where the critical URLLC quality-of-service (QoS) is guaranteed while maximizing the eMBB ergodic capacity. The spatial degrees of freedom at the BS are leveraged to fulfill the URLLC decoding requirements without jeopardizing the performance of the eMBB users. A similar study but for a distributed setup was conducted in~\cite{Abedin2019} where a joint user association and resource allocation problem is formulated for the downlink of a fog network, considering the coexistence of URLLC and eMBB services for \textit{internet-of-things} (IoT) applications. An analytic hierarchy process was proposed for setting the priorities of the services and to formulate a two-sided matching game where a stable association between the fog network infrastructure and IoT devices is established.
\begin{table*}[!t]
\caption{Features of the 5G use cases}
\centering
\begin{tabular}{>{\centering\arraybackslash}p{0.15\textwidth}| p{0.23\textwidth}| p{0.26\textwidth}| p{0.24\textwidth}}
\rowcolor{white!29} & \Centering \textbf{eMBB} & \Centering\textbf{URLLC} & \Centering\textbf{mMTC} \\ \hline
characteristics & high rate, moderate reliability & low latency, ultra reliability, low rate & low  rate, large connectivity \\
traffic & large payload, several devices & small payload, few devices & small payload, massive devices\\
activation pattern & stable & intermittent & intermittent\\
time span & long, multiple resources & short, slot & long, multiple resources \\
frequency span & single/multiple resources  & multiple resources, diversity & single resource \\
scheduling & to prevent access collision & for high reliability & infeaseable \\
random access & if needed & to support intermittency & fundamental \\
target & maximize data rate & meet latency and reliability requirements  & maximize supported arrival rate \\
reliability requirement & \Centering $\sim\!\!\!\!10^{-3}$ & \Centering $\sim\!\!\!\!10^{-5}$ & \Centering $\sim\!\!\!\!10^{-1}$\\
applications & video streaming, augmented reality, entertainment &  connected factories, traffic safety, autonomous vehicles, telemedicine & internet of things, low-power sensors, smart cities
\end{tabular}
\label{tab:features}
\end{table*}

The coexistence between eMBB and URLLC is of most interest~\cite{Matera2018,Alsenwi2019,Abreu2019,Tominaga2021,Saggese2022}, and is mainly  handled with three alternative techniques, herein listed in descending order of complexity: 
\begin{itemize}
\item \textit{successive interference cancellation} (SIC), with which the receiver iteratively decode and remove the contributions of a specific service from the cumulative received signal. This approach requires that the receiver has access to the \textit{channel state information} (CSI) to be able to perform the multi-stage decoding, with decreasing levels of interference, to the required successful decoding probability.
\item \textit{puncturing} (PUNC), consisting in preventing the inter-service interference. In the downlink, whenever the transmitter has to transmit a URLLC signal, then the eMBB signals are dropped over the channel uses involved by th URLLC transmission.  In the uplink, the receiver uses an erasure decoder to discard the eMBB signals, provided that it is able to detect the presence of URLLC transmissions, e.g., via energy detection.
\item \textit{superposition coding} (SPC), with which the transmitter simply sends a linear combination of eMBB and URLLC signals. At the receiver, both for the uplink and the downlink, the inter-service interference is treated as uncorrelated noise (TIN). Again, this approach requires the receiver to be able to detect the presence of the undesired transmissions.  
\end{itemize}
In~\cite{Matera2018} the coexistence of URLLC and eMBB services in the uplink of a C-RAN architecture with shared analog fronthaul links is analyzed, accounting for SIC, puncturing, and TIN. This work provides an information-theoretic study in the performance of URLLC and eMBB traffic under both H-OMA and H-NOMA, by considering standard cellular models with additive Gaussian noise links and a finite inter-cell interference. The main conclusions are that NOMA achieves higher eMBB rates with respect to H-OMA, while guaranteeing reliable low-rate URLLC communication with minimal access latency. Moreover, H-NOMA under SIC is seen to achieve the best performance, while, unlike the case with digital capacity-constrained fronthaul links, TIN always outperforms puncturing. 
A similar analysis is conducted in~\cite{Kassab2019} including both uplink and downlink of C-RAN without analog fronthaul but considering practical aspects, such as fading, the lack of CSI for URLLC transmitters, rate adaptation for eMBB transmitters and finite fronthaul capacity. 
Abreu \textit{et al.} in~\cite{Abreu2019} analyzes both the H-OMA and H-NOMA options for eMBB traffic, and grant-free URLLC in the uplink accounting for minimum mean square error (MMSE) receivers with and without SIC, and under the assumption of Rayleigh fading channels. The resulting outage probability and achievable rates show that TIN is mostly beneficial in sufficiently high-SNR regime when SIC is employed or, in some cases, with low URLLC load. Otherwise, H-OMA supports higher loads for both services simultaneously.
Recently,~\cite{Tominaga2021} proposed an approach to improve the supported loads for URLLC in the uplink, for both H-OMA and H-NOMA in presence of eMBB traffic, showing the superiority of H-NOMA in ensuring the reliability requirements of both the services. A similar analysis but for the downlink is conducted in~\cite{Almekhlafi2021,Saggese2022} where optimal resource allocation strategies and H-NOMA are combined to satisfy the eMBB and URLLC QoS constraints, under the assumption of perfect eMBB CSI and statistical URLLC 
CSI knowledge. \vfill

The information-theoretic framework used by the aforementioned works to characterize the performance achieved by eMBB and URLLC users cannot be applied to massive MIMO scenarios, for different reasons. Establishing the rate (or the spectral efficiency) of the eMBB users in the \textit{ergodic (infinite-blocklength) regime}, upon the block-fading channel model, is sound as the eMBB codewords spans an infinite number of independent fading realizations. Nevertheless, as per the performance of the URLLC users in a quasi-static fading scenario, the use of the outage capacity, whose analysis includes infinite-blocklength assumptions, leads to an inaccurate evaluation of the error probability, as demonstrated in~\cite{Ostman2021}. In addition, outage capacity analyses do not capture the effects of the CSI acquisition overhead when pilots are used to estimate the uplink channel. As an alternative, finite-blocklength analyses have been proposed for URLLC in conventional cellular networks~\cite{Almekhlafi2021,Saggese2022}, co-located massive MIMO networks~\cite{ZengJ2020,Ren2020} and cell-free massive MIMO networks~\cite{Nasir2021}, and rely on the information-theoretic bounds and tools developed in~\cite{Polyanskiy2010}, e.g., the well known \textit{normal approximation}. However, the work in~\cite{Ostman2021} proved that the normal approximation is not accurate  in the region of low error probabilities of interest in URLLC ($<\!10^{-4}$), especially as the number of antennas at the BS increases, and in presence of imperfect CSI. Importantly, \"{O}stman \textit{et al.} in~\cite{Ostman2021} provided a more rigorous finite-blocklength information-theoretic framework relying on the use of a mismatched decoding~\cite{Scarlett2014}, and of the \textit{saddlepoint approximation}~\cite{Martinez2011} for evaluating the error probability of the URLLC users in co-located massive MIMO systems. This framework, priory developed for wireless fading channels in~\cite{Yang2014,Durisi2016,Ostman2019}, accounts for linear signal processing, imperfect CSI and instantaneous channel estimation error, and additive uncorrelated noise including multi-user interference. However, the analysis of~\cite{Ostman2021} is limited to the URLLC regime, and the coexistence with the eMBB is yet to be investigated under a unified information-theoretic framework.       

\subsection{CONTRIBUTIONS}
Our contributions can be summarized as follows.
\begin{itemize}
\item We investigate the non-orthogonal multiplexing of the eMBB and the URLLC, in the downlink of a multi-cell massive MIMO system, by providing a unified information-theoretic framework that combines an infinite-blocklength analysis to assess the SE of the eMBB and
a finite-blocklength analysis to assess the error probability of the URLLC.
\item Unlike prior works wherein the URLLC performance is inappropriately evaluated by the use of the outage capacity analysis or the error probability obtained  via the normal approximation, in this work the finite-blocklength information-theoretic analysis relies on the results and tools established in~\cite{Ostman2021}, where mismatched receivers and saddlepoint approximation are assumed, but the coexistence between and URLLC and eMBB was not investigated. 
\item The proposed unified framework accommodates two alternative coexistence strategies: PUNC and SPC. The former prevents the inter-service interference to protect the URLLC reliability, whereas the latter accepts it to maintain the eMBB service. In addition, the analytical framework accounts for imperfect CSI acquisition at the BSs via uplink pilot transmissions, pilot contamination and pilot overhead, spatially correlated channels and the lack of CSI at the users. 
\item We numerically evaluate the performance achieved by PUNC and SPC under different precoding schemes, namely maximum ratio, regularized zero-forcing and multi-cell MMSE, and different power allocation strategies, i.e., equal power allocation, weighted fractional power allocation and optimal power allocation maximizing the product SINR throughout the network. The coexistence between eMBB and URLLC is explored in various scenarios, including different configurations of the time-division duplex radio frame, and different URLLC random activation patterns. 
\item The results of our comprehensive simulation campaign highlight the clear superiority of SPC over PUNC in most of the considered operating regimes. The main limitation of SPC, namely the caused multi-user interference, is often overcome by using regularized zero-forcing and multi-cell MMSE, which in turn hinge on a high-quality CSI acquisition. Whenever these precoding techniques cannot be implemented due to complexity or hardware constraints, the URLLC reliability requirements can be met by fine-tuning the parameters of the proposed weighted fractional power allocation. Conversely, performing PUNC is necessary to preserve the URLLC performance if the interference cancellation via precoding is ineffective, for instance, when pilot contamination is high or the multi-user interference is excessive.
\item Pilot contamination among URLLC users is particularly destructive. This led us to devise a pilot assignment policy that prioritizes the URLLC users. In our approach, we primarily assign unique orthogonal pilots to the URLLC users, admitting pilot reuse only among eMBB users. If doable, orthogonal pilots are assigned within cells to prevent the intra-cell pilot contamination, and if the uplink training length is sufficiently large, then mutually orthogonal pilots are guaranteed to everyone.    
\end{itemize}

\subsection{PAPER OUTLINE}
The remainder of this paper is organized as follows. In~\Secref{sec:system-model}, we introduce the system model of the multi-cell massive MIMO system, including the description of the uplink training and a unified framework for the data transmission stage accounting for both puncturing and superposition coding techniques. In~\Secref{sec:performance-analysis} we present the information-theoretic analyses in the infinite-blocklength regime and finite-blocklength regime for the eMBB and the URLLC performance evaluation, respectively. \Secref{sec:precoding-power-control} details the precoding techniques and power allocation strategies to deal with the coexistence of eMBB and URLLC users. Simulation results and discussions are provided in~\Secref{sec:simulation-results}, while the main findings of this work are discussed in~\Secref{sec:conclusion}.

\subsection{NOTATION}
Vectors and matrices are denoted by boldface lowercase and boldface uppercase letters, respectively. Calligraphy uppercase letters denote sets, while $\mathbb{C}$ and $\mathbb{R}$ represent the sets of complex and real numbers, respectively.
$\EX{\cdot}$ indicates the expectation operator, while $\Prx{\cdot}$ denotes the probability of a set. $x^+$ represents the \textit{positive part} function, namely $x^+\!=\!\max\{x,0\}$, and $\lfloor \cdot \rfloor$ denotes the \textit{floor} function. The natural logarithm is indicated by $\log(\cdot)$ and $Q(\cdot)$ describes the Gaussian \textit{Q}-function.
$\CG{\boldsymbol{\mu}}{\boldsymbol{\Sigma}}$ describes a circularly symmetric complex Gaussian distribution with mean $\boldsymbol{\mu}$ and covariance matrix $\boldsymbol{\Sigma}$.  
The superscripts $(\cdot)\trans$, $(\cdot)^{\ast}$ and $(\cdot)\herm$ denote the transpose, the conjugate and the conjugate transpose (Hermitian) operators, respectively.
$\tr(\bA)$ indicates the trace of the matrix $\bA$, while $\norm{\ba}$ denotes the $\ell_2$-norm of the vector $\ba$. The notation $[\bA]_{:,i}$ indicates the $i$th column of the matrix $\bA$. 
$\bI_N$ represents the identity matrix of size $N\!\times\!N$. \Tableref{tab:tab1} introduces the notation definition used in the system model of this paper.   

\begin{table}[!t]
\caption{System Model Notation}
\setlength{\tabcolsep}{2pt}
\begin{tabular}{|>{\centering\arraybackslash}p{25pt}|p{81pt}|>{\centering\arraybackslash}p{25pt}|p{95pt}|}
\hline
Symbol& 
Description&
Symbol& 
Description\\
\hline
$L$&n. of cells &$K$ &n. of users/cell \\
$M$& n. of BS antennas &$\Ku$ &n. of URLLC users/cell \\
$\alpha$& $\Ku/K \in (0,1)$ &$\Ke$ &n. of eMBB users/cell \\
$\tc$&TDD frame length &$\setKu_j$ &URLLC users set in cell $j$ \\
$\tp$&UL training length &$\setKe_j$ &eMBB users set in cell $j$ \\
$\td$&DL data trans. length & $T$ & n. of slots in a TDD frame \\
$\bh^{j}_{lk}$& \multicolumn{3}{l|}{channel between BS $j$ and user $k$ in cell $l$, vector $\mC^M$} \\
$\hhat^{j}_{lk}$ &estimate of $\bh^{j}_{lk}$ &$\htilde^{j}_{lk}$ &estimation error $\bh^{j}_{lk}\!-\!\hhat^{j}_{lk}$\\
$\bR^{j}_{lk}$ &correl. matrix of $\bh^{j}_{lk}$ &$\beta^{j}_{lk}$ &average channel gain of $\hhat^{j}_{lk}$ \\
$\bC^{j}_{lk}$ &correl. matrix of $\htilde^{j}_{lk}$ &$f$ & pilot reuse factor \\
$p\pilot_{jk}$& UL pilot power &$\rhoMax_j$  & max transmit power at BS $j$   \\
$\setP_{jk}$ &\multicolumn{3}{l|}{set of all the users using the same pilot as user $k$ in cell $j$} \\ 
$A^t_{jk}$ &\multicolumn{3}{l|}{1 if URLLC user $k$ in cell $j$ is active in slot $t$, 0 otherwise} \\
$\au$ &\multicolumn{3}{l|}{parameter of the Bernoulli distribution that draws $A^t_{jk}$} \\
$\datae_{jk}[n]$ &\multicolumn{3}{l|}{data transmitted by BS $j$ to eMBB user $k$ in channel use $n$} \\
$\datau_{ji}[n]$ &\multicolumn{3}{l|}{data transmitted by BS $j$ to URLLC user $i$ in channel use $n$} \\
$\bw_{jk}$ &\multicolumn{3}{l|}{precoding vector, $\mC^M$, used by BS $j$ to its user $k$} \\
$\nvu$& UL noise variance &$\rhou_{ji}$& DL power to URLLC user $i$ \\
$\nvd$& DL noise variance & $\rhoe_{jk}$& DL power to eMBB user $k$ \\
$g^{li}_{jk}$ &\multicolumn{3}{l|}{precoded DL channel from BS $l$ using $\bw_{li}$ to user $k$ in cell $j$}\\
$\widehat{g}^{li}_{jk}$& estimate of $g^{li}_{jk}$  &$\nd$ &URLLC codeword length \\
$\epsilon^{\mathsf{dl}}_{jk}$& DL error probability &$\eta^{\mathsf{dl}}$ &DL network availability \\
$\nu$ &\multicolumn{3}{l|}{exponent characterizing the fractional power allocation (FPA)} \\
$\omega$ &\multicolumn{3}{l|}{FPA weight tuning the power allocated to the URLLC users} \\
\hline
\end{tabular}
\label{tab:tab1}
\end{table}

\section{SYSTEM MODEL}
\label{sec:system-model}

Let us consider a multi-cell massive MIMO system with $L$ cells, each one served by a BS that is placed at the cell-center and equipped with $M$ co-located antennas. Each cell covers a square area of $D\!\times\!D$ km$^2$, and provide service to $K$ users. It holds that $M\!\gg\!K$ so that interference suppression can be efficiently carried out by exploiting the spatial degrees of freedom. A fraction $0\!\leq\!\alpha\!\leq\!1$ of the $K$ users requests a URLLC service, e.g., a vehicle in \textit{cellular vehicle-to-everything} (C-V2X) use cases for intelligent transportation systems, or a machine in factory automation use cases for ``Industry 4.0''. Letting $\Ku\!=\!\alpha K$ be the number of URLLC users per cell, then $\Ke\!=\!K\!-\!\Ku$ is the number of eMBB users per cell. 
The set including the indices of the eMBB and URLLC users in cell $j$ is denoted as $\setKe_j$ and $\setKu_j$, respectively. 

\subsection{TDD PROTOCOL AND FRAME STRUCTURE}
The considered system operates in time-division duplex (TDD) mode to facilitate CSI acquisition and limit the estimation overhead. In addition, we assume that the channel is reciprocal as a result of a perfect calibration of the RF chains. By leveraging the channel reciprocity, the channel estimates acquired by the BS in the uplink are then utilized in the downlink to design the transmit precoding vectors. As channel hardening holds for co-located massive MIMO systems with sufficiently large antenna arrays in most of the propagation environments, we assume that the users do not estimate the downlink channels, and reliably decode downlink data solely relying on the knowledge of the statistical CSI. Hence, the TDD protocol consists of three phases: $(i)$ pilot-based uplink training, $(ii)$ uplink data transmission, and $(iii)$ downlink data transmission. 

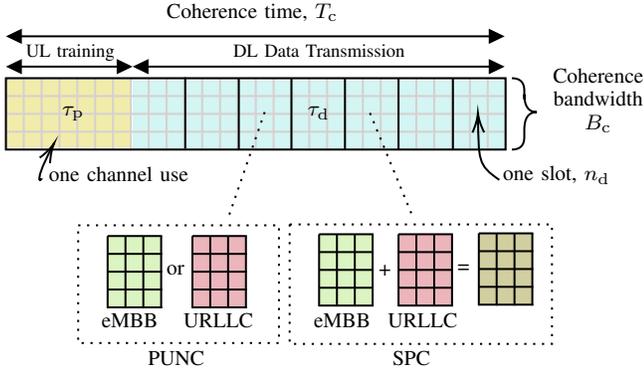
\begin{figure}
\input{TDDframe.tex}
\caption{An illustration of the TDD frame assuming no uplink data transmission phase, and representing the resource allocation in case of puncturing (PUNC) and superposition coding (SPC) operation.}
\label{fig:TDDframe}
\end{figure}

The time-frequency resources are structured in TDD frames, each one grouping a set of subcarriers and time samples over which the channel response is assumed being frequency-flat and time-invariant. The TDD frame must accommodate the aforementioned protocol phases and supporting all the users, thus its size is designed to match that of the smallest user's coherence block in the network. 
As shown in~\Figref{fig:TDDframe}, the TDD frame consists of $\tc \!=\! \Tc \Bc$ samples (or \textit{channel uses}) where $\Tc$ is the coherence time and $\Bc$ is the coherence bandwidth. $\tp$ channel uses out of $\tc$ are spent for the uplink CSI acquisition, whereas the remaining channel uses are devoted to the uplink and downlink data transmission. Since, in this paper, we only focus on the downlink operation, we assume that $\td \!=\! \tc\!-\!\tp$ is the length of the downlink data transmission phase, without loss of generality. The latter is divided in $T$ slots of equal length. As conventionally assumed in the \textit{ergodic regime}, an eMBB transmission spans multiple (theoretically an infinite number of) TDD frames, wherein the channel realizations evolve independently according to the block-fading model. To evaluate the spectral efficiency achieved by the eMBB users, we look at a single TDD frame and resort to the information-theoretic bounds and tools in the infinite-blocklength regime~\cite{redbook,massivemimobook}.
Whereas, URLLC transmissions are confined in time to meet the very strict latency requirements and are allowed to span only one slot. Hence, the number of channel uses in a slot equals the URLLC codeword length.
We assume a random activation pattern of the URLLC users. Within a TDD frame, a URLLC user may be active in multiple slots. To characterize the error probability of the URLLC transmissions, we look separately at each single slot of a TDD frame and resort to the finite-blocklength information-theoretic bounds and tools presented in~\cite{Ostman2021}.     

\subsection{CHANNEL MODEL AND UPLINK TRAINING}
The channel response between the $k$-th user in cell $l$ and the BS in cell $j$ is denoted by the $M$-dimensional complex-valued vector $\bh^j_{lk}$. We assume correlated Rayleigh fading channels, that is $\bh^j_{lk}\!\sim\!\CG{\bzero_M}{ \bR^j_{lk}}$, where $\bR^j_{lk}\! \in\! \mC^{M \times M}$ is the positive semi-definite spatial correlation matrix. The corresponding average channel gain (or large-scale fading coefficient) is given by $\beta^j_{lk}\! =\! \tr(\bR^j_{lk})/M$. Large-scale fading quantities are assumed to be known at the BS.

In the uplink training phase, each user transmits a pilot sequence that spans $\tp$ channel uses. The pilot sequence of user $k$ in cell $j$ is denoted by $\bphi_{jk} \in \mC^{\tp}$. All the pilot sequences are drawn from a set of $\tp$ mutually orthogonal pilots, thereby the inner product between two pilots equals either $\tp$ if the sequences are identical or $0$ if they are mutually orthogonal. Notice that re-using the pilots throughout the network might be unavoidable as the share of the TDD frame reserved to the training is limited and, importantly, as the CSI acquisition overhead significantly degrades the spectral efficiency. Pilot reuse gives rise to additional interference, known as \textit{pilot contamination}~\cite{Marzetta2010}, that degrades the quality of the acquired CSI and correlates the channel estimates.   
The cumulative uplink signal received at BS $j$, denoted by $\bY\pilot_j \in \mathbb{C}^{M\times\tp}$, reads
\begin{align}
\label{eq:cell:pilot-matrix}
\bY\pilot_j = \sum\limits^{K}_{k=1}\sqrt{p\pilot_{jk}}\bh^j_{jk}\bphi\trans_{jk} + \sum  \limits^L_{\substack{l=1 \\ l\neq j}}\sum\limits^K_{i=1} \sqrt{p\pilot_{li}} \bh^j_{li}\bphi\trans_{li} + \bN\pilot_j \ ,  
\end{align} 
where $p\pilot_{jk}$ is the transmit pilot power, and $\bN\pilot_j$ is the additive receiver noise with \iid~elements distributed as $\CG{0}{\nvu}$, with $	\nvu$ being the receiver noise variance in the uplink.
To estimate the channel of user $k$ in its own cell, $\bh^j_{jk}$, BS $j$ correlates $\bY\pilot_j$ with the known pilot sequence $\bphi_{jk}$ as
\begin{align}
\label{eq:pilot-despreading}
\by\pilot_{jjk} \!&=\! \bY\pilot_j \bphi\conj_{jk} \nonumber \\
&=\! \sqrt{p\pilot_{jk}} \tp \bh^j_{jk} + \sum\limits^{K}_{\substack{i=1 \\ i \neq k}} \sqrt{p\pilot_{ji}} \bh^j_{ji} \bphi\trans_{ji} \bphi\conj_{jk} \nonumber \\
&\quad+ \sum\limits^L_{\substack{l=1 \\ l\neq j}} \sum\limits^{K}_{i=1} \sqrt{p\pilot_{li}} \bh^j_{li} \bphi\trans_{li} \bphi\conj_{jk} + \bN\pilot_j \bphi\conj_{jk} \ .    
\end{align}
In~\eqref{eq:pilot-despreading}, the second term of the rightmost right-hand side represents the intra-cell pilot contamination term, while the third term quantifies the inter-cell pilot contamination. 
A conventional pilot allocation strategy consists in assigning mutually orthogonal pilots to users within the same cell, and re-using the pilot sequences over different cells~\cite{massivemimobook}. This is a reasonable choice as intra-cell pilot contamination is presumably stronger than inter-cell pilot contamination. We let $\tp\!=\!fK$ where $f$ is referred to as \textit{pilot reuse factor}. 
Importantly, in order not to jeopardize the ultra-reliability of the URLLC transmissions, we assume that unique orthogonal pilot sequences are assigned to all the URLLC users in the network, if doable (namely when $\tp\!>\!L\Ke$). Summarizing, the pilot allocation strategy we propose primarily aims to prevent URLLC users from being affected of pilot contamination, and secondarily to prevent intra-cell pilot contamination. Finally, if $\tp$ is sufficiently large, that is $\tp\!\geq\!LK$, then mutually orthogonal pilots can be guaranteed to everyone. Let us define the set
\begin{align}
\label{eq:set-copilot-users}
\setP_{jk} \!=\! \left\{(l,i): \bphi_{li} \!=\! \bphi_{jk}, \, l\!=\!1,\ldots,L, \, i\!=\!1,\ldots,K\right\} \ ,
\end{align} 
including the indices of all the users (and of the corresponding cells) that use the same pilot as user $k$ in cell $j$. Hence, we can rewrite~\eqref{eq:pilot-despreading} as 
\begin{align}
\label{eq:pilot-despreading-final}
\by\pilot_{jjk} \!=\! \sqrt{p\pilot_{jk}} \tp \bh^j_{jk} + \tp \!\! \sum\limits_{(l,i) \in \mathcal{P}_{jk}\setminus(j,k)} \!\!\! \sqrt{p\pilot_{li}}  \bh^j_{li} + \bN\pilot_j \bphi\conj_{jk}.    
\end{align}
The processed uplink signal, $\by\pilot_{jjk}$, is a \textit{sufficient statistic} for the estimation of $\bh^j_{jk}$. Upon the knowledge of the spatial correlation matrices, BS $j$ can compute the minimum mean-squared error (MMSE) estimate of $\bh^j_{jk}$, denoted by $\hhat^j_{jk}$, based on the observation $\by\pilot_{jjk}$ as~\cite{massivemimobook}
\begin{align}
\label{eq:MMSE-estimate}
\hhat^j_{jk} \!&=\! \sqrt{p\pilot_{jk}} \bR^j_{jk} \bPsi^j_{jk} \by\pilot_{jjk} \,
\end{align}
where 
\begin{align}
\label{eq:MMSE-estimate-Psi}
\bPsi^j_{jk} \!=\! {\left(\sum\limits_{(l,i) \in \mathcal{P}_{jk}} p\pilot_{li} \tp \bR^j_{li} + \sigma^2_{\mathrm{ul}}\bI_{M_j}\right)}\inv.  
\end{align}   
The estimation error is given by $\htilde^j_{jk} \!=\! {\bh}^j_{jk}\!-\! \hhat^j_{jk}$, and has correlation matrix $$\bC^j_{jk}\!=\! \EX{\htilde^j_{jk}(\htilde^j_{jk})\herm} = \bR^j_{jk}\!-\!p\pilot_{jk}\tp\bR^j_{jk}\bPsi^j_{jk}\bR^j_{jk}.$$ It follows that $\widetilde{\bh}^j_{jk}$ and $\widehat{\bh}^j_{jk}$ are independent random variables distributed as
\begin{align*}
\widetilde{\bh}^j_{jk}&\sim\CG{\bzero_M}{\bC^j_{jk}}\ , \\
\widehat{\bh}^j_{jk}&\sim\CG{\bzero_M}{\bR^j_{jk}\!-\!\bC^j_{jk}} \ .
\end{align*}

\subsection{DOWNLINK TRANSMISSION}
In the downlink transmission phase, each BS transmits payload data to all the active users of its cell.
Let $A^t_{jk}$ be a coefficient that equals 1 if a URLLC transmission takes place at the $t$-th slot for URLLC user $k$ in cell $j$, and 0 otherwise. This coefficient models the random activation pattern of the URLLC users which follows a Bernoulli distribution with parameter $\au$, $A^t_{jk} \!\sim\!\mathcal{B}\mathit{ern}(\au)$.
To handle the coexistence of eMBB and URLLC users in the downlink, we consider two transmission techniques: $(i)$ puncturing, and $(ii)$ superposition coding.
Under puncturing, whenever a URLLC transmission is triggered by a BS in a certain slot, all the eMBB transmissions therein are dropped. However, the eMBB service can be still guaranteed in the remaining slots of the frame where no URLLC users are active. Under superposition coding, eMBB transmissions occur in all the slots and each BS linearly combines eMBB and URLLC signals whenever
URLLC transmissions are triggered. 

The analytical framework detailed next is generalized, namely holds for both the aforementioned transmission techniques upon setting, for an arbitrary BS $j$ and slot $t$, the coefficient
\begin{align*}
\Atj = \begin{cases}
{\Bigg(1-\sum\limits_{i \in \setKu_j} A^t_{ji} \Bigg)}^{\!\!+}, \quad& \text{for puncturing}, \\
1 \ , \quad& \text{for superposition coding}. 
\end{cases}
\end{align*}
Let $\datae_{jk}[n]$ or $\datau_{jk}[n]$ be the data symbol transmitted by BS $j$ to user $k$ over an arbitrary channel use $n$, if $k$ is an eMBB user or a URLLC user, respectively.
We assume that $\datas_{jk}[n] \!\sim\! \CG{0}{1}$, with $\mathsf{s} = \{\mathsf{e},\mathsf{u}\}$.    
A slot consists of $\nd$ channel uses, with $\nd \!=\! \lfloor \td/T \rfloor$, and equals the length of the URLLC codeword.
The data symbol is precoded by using the $M$-dimensional precoding vector $\bw_{jk}$, which is function of the CSI acquired at the BS during the uplink training. It also holds $\EX{\norm{\bw_{jk}}^2}=1$. The data signal transmitted by BS $j$ over an arbitrary channel use $n$ of slot $t$ is given by
\begin{align}
\label{eq:transmitted-data-signal:punc}
\bx^t_j[n] = \Atj\!\sum\limits_{k \in \mathcal{K}^{\mathsf{e}}_j} \! \sqrt{\rhoe_{jk}} \bw_{jk}\datae_{jk}[n] \!+\!\! \sum\limits_{i \in \setKu_j} \! A^t_{ji} \sqrt{\rhou_{ji}} \bw_{ji}\datau_{ji}[n],   
\end{align}
with $n=1,\ldots,\nd$, and where $\rhoe_{jk}$ and $\rhou_{ji}$ are the downlink transmit powers used by BS $j$ to its eMBB user $k$ and URLLC user $i$, respectively, satisfying the following per-BS power constraint
\begin{align}
\EX{\norm{\bx^t_j[n]}^2} = \Atj\sum\limits_{k \in \mathcal{K}^{\mathsf{e}}_j} {\rhoe_{jk}} \!+\! \sum\limits_{i \in \setKu_j} A^t_{ji} {\rhou_{ji}} \leq \rhoMax_j \ ,    
\label{eq:data-power-constraint}
\end{align}
with $j \!=\! 1,\ldots,L$, and where $\rhoMax_j$ is the maximum transmit power at BS $j$.
The data signal received at user $k$ in cell $j$ over an arbitrary channel use $n$ of slot $t$ is denoted as $y^{t,\mathsf{s}}_{jk}[n]$, with $\mathsf{s} = \{\mathsf{e},\mathsf{u}\}$. 
In line with the conventional massive MIMO operation, we assume that the users do not acquire the instantaneous downlink CSI, but rather rely on a mean value approximation of their downlink precoded channels. Such approximation is accurate if channel hardening occurs.
If user $k$ in cell $j$ is an eMBB user, namely $k \in \mathcal{K}^{\mathsf{e}}_j$, then its received data signal over an arbitrary channel use $n$ of slot $t$ can be written as in~\eqref{eq:received-data-signal-eMBB} at the top of the next page, where $w_{jk}[n]\!\sim\!\CG{0}{\nvd}$ is the \iid~receiver noise with variance $\nvd$, and we have defined $g^{li}_{jk} \!=\! (\bh^l_{jk})\herm\bw_{li}$, namely the precoded downlink (scalar) channel between the BS in cell $l$, using the precoding vector intended for its user $i$, and the $k$-th user in cell $j$. 
%
\begin{figure*}[!t]
\normalsize
\setcounter{counter_1}{\value{equation}}
\setcounter{equation}{8}
\begin{align}
\label{eq:received-data-signal-eMBB}
y^{t,{\mathsf{e}}}_{jk}[n] &= \underbrace{\vphantom{\sum\limits_{i \in \mathcal{K}^{\mathsf{e}}_j}}\EX{g^{jk}_{jk}}\Atj\sqrt{\rhoe_{jk}}\datae_{jk}[n]}_{\text{desired signal}} + \underbrace{\vphantom{\sum\limits_{i \in \mathcal{K}^{\mathsf{e}}_j}}\left(g^{jk}_{jk}\!-\!\EX{g^{jk}_{jk}}\right)\Atj\sqrt{\rhoe_{jk}}\datae_{jk}[n]}_{\text{self-interference}} +\! \underbrace{\sum\limits_{i \in \setKu_j} g^{ji}_{jk} A^t_{ji} \sqrt{\rhou_{ji}}  \datau_{ji}[n]}_{\text{intra-cell inter-service interference}}  \nonumber \\
&\quad+\!\!\! \underbrace{\vphantom{\sum\limits^L_{\substack{l=1 \\ l\neq j}}}\sum\limits_{i \in \mathcal{K}^{\mathsf{e}}_j\setminus\{k\}} g^{ji}_{jk}\Atj\sqrt{\rhoe_{ji}}  \datae_{ji}[n]}_{\text{intra-cell intra-service interference}} + \underbrace{\sum\limits^L_{\substack{l=1 \\ l\neq j}} \sum\limits_{i \in \mathcal{K}^{\mathsf{e}}_l} g^{li}_{jk} \Atl \sqrt{\rhoe_{li}}  \datae_{li}[n]}_{\text{inter-cell intra-service interference}}
+ \underbrace{\sum\limits^L_{\substack{l=1 \\ l\neq j}} \sum\limits_{i \in \setKu_l} g^{li}_{jk} A^t_{li} \sqrt{\rhou_{li}} \datau_{li}[n]}_{\text{inter-cell inter-service interference}}
+ \underbrace{\vphantom{\sum\limits^L_{\substack{l=1 \\ l\neq j}}}w_{jk}[n]}_{\text{noise}}
\end{align}
\setcounter{equation}{\value{equation}}
\hrulefill
\vspace*{-3mm}
\end{figure*}
%
If user $k$ in cell $j$ is a URLLC user, its received data signal over an arbitrary channel use $n$ in slot $t$ can be written as in~\eqref{eq:received-data-signal-URLLC} at the top of the next page. \Eqref{eq:received-data-signal-eMBB} emphasizes the fact that user $k$ in cell $j$ solely knows the statistical CSI of the downlink channel, that is $\EX{g^{jk}_{jk}}$. The second term in~\eqref{eq:received-data-signal-eMBB} represents the self-interference due to this lack of instantaneous CSI, referred to as \textit{beamforming gain uncertainty}. Going forward, the intra-cell inter-service interference and intra-cell intra-service interference terms represent the interference caused by the URLLC and eMBB users of cell $j$, respectively. This is presumably stronger than the inter-cell interference caused by the eMBB users (i.e., intra-service) and the URLLC users (i.e., inter-service) in the other cells. A similar distinction of the various signal contributions is reported in~\eqref{eq:received-data-signal-URLLC} for URLLC user $k$ in cell $j$. In this case, the lack of instantaneous CSI at the user will be highlighted in the next section.     

\begin{figure*}[!t]
\normalsize
\setcounter{counter_1}{\value{equation}}
\setcounter{equation}{9}
\begin{align}
\label{eq:received-data-signal-URLLC}
y^{t,{\mathsf{u}}}_{jk}[n] \!=\! \underbrace{\vphantom{\sum\limits^L_{\substack{l=1 \\ l\neq j}}} g^{jk}_{jk}A^t_{jk}\sqrt{\rhou_{jk}}\datau_{jk}[n]}_{\text{desired signal}} 
+\!\!\!\! \underbrace{\vphantom{\sum\limits^L_{\substack{l=1 \\ l\neq j}}} \sum\limits_{i \in \setKu_j\setminus\{k\}} \!\! g^{ji}_{jk} A^t_{ji} \sqrt{\rhou_{ji}} \datau_{ji}[n]}_{\text{intra-cell intra-service interference}}
+\! \underbrace{\sum\limits^L_{\substack{l=1 \\ l\neq j}} \sum\limits_{i \in \setKu_l} g^{li}_{jk}  A^t_{li} \sqrt{\rhou_{li}} \datau_{li}[n]}_{\text{inter-cell intra-service interference}}
\!+\!\underbrace{\vphantom{\sum\limits^L_{\substack{l=1 \\ l\neq j}}} \sum\limits^L_{\substack{l=1}} \sum\limits_{i \in \mathcal{K}^{\mathsf{e}}_l}g^{li}_{jk} \Atl\sqrt{\rhoe_{li}} \datae_{li}[n]}_{\text{inter-service interference}}
+ \underbrace{\vphantom{\sum\limits^L_{\substack{l=1 \\ l\neq j}}}w_{jk}[n]}_{\text{noise}}
\end{align}  
\setcounter{equation}{\value{equation}}
\hrulefill
\vspace*{4pt}
\end{figure*}
 
\section{PERFORMANCE ANALYSIS}
\label{sec:performance-analysis}
In this section, we evaluate the downlink performance of eMBB and URLLC users.
As per the eMBB users, we consider the spectral efficiency (SE) by applying the infinite-blocklength information-theoretic results established in the \textit{ergodic regime}~\cite{Tse2005,redbook,massivemimobook}.  
An achievable downlink SE, namely a lower-bound on the ergodic downlink capacity, can be obtained by applying the popular \textit{hardening bound} technique~\cite{redbook,massivemimobook} on the signal model in~\eqref{eq:received-data-signal-eMBB}, by treating all the interference sources as uncorrelated noise. Specifically, an achievable downlink spectral efficiency of an arbitrary eMBB user $k$ in cell $j$, is given by 
\begin{align}
\label{eq:eMBB:SE}
\mathsf{SE}^{\mathsf{e}}_{jk} = \dfrac{\td}{\tc} \dfrac{1}{T}\sum^T_{t=1} \log_2(1+\mathsf{SINR}^{t,\mathsf{e}}_{jk}), \text{ [bits/s/Hz]} \ ,
\end{align}
where $\td/\tc$ accounts for the estimation overhead, 
\begin{align} \label{eq:eMBB:SINR}
\mathsf{SINR}^{t,\mathsf{e}}_{jk} &= \frac{\Atj\rhoe_{jk}\left|\EX{g^{jk}_{jk}}\right|^2}{\sum\limits^L_{l=1} \sum\limits^K_{i=1}\varrho^t_{li}\EX{|g^{li}_{jk}|^2}\!-\!\Atj\rhoe_{jk}\left|\EX{g^{jk}_{jk}}\right|^2\!\!+\!\nvd} \ ,
\end{align}
is the effective SINR of user $k \in \mathcal{K}^{\mathsf{e}}_j$, where the expectations are taken with respect to the random channel realizations, and
\begin{align}
\label{eq:varrho-def}
\varrho^t_{li} = \begin{cases} 
 A^t_{li} \rhou_{li},&\quad \text{if } i \in \setKu_l \ ,\\
 \Atl{\rhoe_{li}},&\quad \text{if } i \in \mathcal{K}^{\mathsf{e}}_l \ .
\end{cases}
\end{align}
The expression of the achievable SE shown in~\eqref{eq:eMBB:SE} holds for any choice of precoding scheme, any channel estimator and any channel distributions. Importantly, it accounts for any choice of coexistence technique between heterogeneous services, namely puncturing or superposition coding.
The infinite-blocklength analysis above is established upon the assumption of block-fading channel model, entailing that each eMBB codeword has infinite length that spans a large number of independent fading realizations. This assumption cannot be applied to the URLLC case. 
As per the URLLC user, we consider a nonasymptotic analysis of the
downlink error probability on a slot basis by applying the finite-blocklength information-theoretic results established in~\cite{Ostman2021}. 
Firstly, we rewrite~\eqref{eq:received-data-signal-URLLC} as
\begin{align}
\label{eq:simplified_channel}
y^{t,{\mathsf{u}}}_{jk}[n] = g^{jk}_{jk} q_{jk}[n]  + z_{jk}[n], \quad n=1,\dots,\nd,
\end{align} 
where $q_{jk}[n] \!=\!A^t_{jk}\sqrt{\rhou_{jk}} \datau_{jk}[n]$, and 
\begin{align}
z_{jk}[n] &= \sum\limits_{i \in \setKu_j\setminus\{k\}} g^{ji}_{jk} q_{ji}[n] +\! \sum\limits_{i \in \mathcal{K}^{\mathsf{e}}_j} g^{ji}_{jk}\Atj\sqrt{\rhoe_{ji}}  \datae_{ji}[n] \nonumber \\ 
&\quad+\sum\limits^L_{\substack{l=1 \\ l\neq j}} \left(\sum\limits_{i \in \setKu_l} g^{li}_{jk} q_{li}[n] +\!\sum\limits_{i \in \mathcal{K}^{\mathsf{e}}_l} g^{li}_{jk}\Atl\sqrt{\rhoe_{li}}  \datae_{li}[n] \right) \nonumber \\
&\quad+ w_{jk}[n] \ . 
\end{align} 
However, URLLC user $k$ in cell $j$ has not access to $g^{jk}_{jk}$, but performs data decoding by only leveraging its mean value, $\widehat{g}^{jk}_{jk} \!=\! \EX{(\bh^j_{jk})\herm\bw_{jk}}$, which is treated as perfect. This estimate is accurate if channel hardening holds. 
Notice that, the precoded channel $g^{jk}_{jk}$ is frequency-flat and time-invariant over the transmission of the $\nd$-length URLLC codeword in slot $t$. Moreover, $g^{jk}_{jk}$ remains constant for any other transmission from BS $j$ to user $k$ over slots in the same TDD frame.  
Given all channels and precoding vectors, the effective noise terms $\{z_{jk}[n]\in\mathbb{C}; n=1,\ldots,\nd\}$ are random variables conditionally \iid~with variance $\sigma_{jk}^2$, i.e., $\CG{0}{\sigma_{jk}^2}$, given by 
\begin{align}
\sigma_{jk}^2 &= \sum\limits_{i \in \setKu_j\setminus\{k\}} A^t_{ji} \rhou_{ji} |g^{ji}_{jk}|^2 +\! \sum\limits_{i \in \mathcal{K}^{\mathsf{e}}_j} \Atj \rhoe_{ji} |g^{ji}_{jk}|^2  \nonumber \\
&\quad+ \! \sum\limits^L_{\substack{l=1 \\ l\neq j}} \left( \sum\limits_{i \in \setKu_l} A^t_{li} \rhou_{li} |g^{li}_{jk}|^2 \!+\!\sum\limits_{i \in \mathcal{K}^{\mathsf{e}}_l} \Atl \rhoe_{li} |g^{li}_{jk}|^2 \right) + \nvd \ . 
\end{align}
To determine the transmitted codeword 
$$\bq_{jk}=[q_{jk}[1],\ldots,q_{jk}[\nd]]\trans \ ,$$ 
user $k$ in cell $j$ employs a \textit{mismatched scaled nearest-neighbor} (SNN) decoder~\cite{Lapidoth2002}, with which selects the codeword $\widetilde{\bq}_{jk}$ from the codebook $\mathcal {C}$ by applying the rule 
\begin{equation}\label{eq:mismatched_snn_decoder}
\widehat{\bq}_{jk}=\argmin_{\widetilde{\bq}_{jk}\in\mathcal{C}} \norm{\by^{t,{\mathsf{u}}}_{jk}-\widehat{g}^{jk}_{jk}\widetilde{\bq}_{jk}}^2 \ ,
\end{equation}
where $\by^{t,{\mathsf{u}}}_{jk} \!=\! [y^{t,{\mathsf{u}}}_{jk}[1],\dots,y^{t,{\mathsf{u}}}_{jk}[\nd]]\trans\!\in\!\mathbb{C}^{\nd}$ is the received data vector.

Let $\epsilon^{\mathsf{dl}}_{jk}=\Prx{\widehat{\bq}_{jk}\neq \bq_{jk}}$ be the downlink error probability experienced by the URLLC user $k$ in cell $j$ achieved by the SNN decoding. An upper bound on $\epsilon^{\mathsf{dl}}_{jk}$ is  obtained by using the standard \textit{random-coding} approach~\cite{gallager1968},
\begin{align}
\label{eq:rcus_fading}
\epsilon^{\mathsf{dl}}_{jk} \!\leq\! \EXs{g^{jk}_{jk}}{\!\Prx{\sum_{n=1}^{\nd} {\imath_s(q_{jk}[n],\by^{t,{\mathsf{u}}}_{jk}[n])} \leq \log\frac{m\!-\!1}{r} \bigg| g^{jk}_{jk}}\!\!}, 
\end{align}
where $m \!=\! 2^b$ is the number of codewords with length $\nd$ that convey $b$ information bits, $r$ is a random variable uniformly distributed in the interval $[0,1]$ and $\imath_s(q_{jk}[n],\by^{t,{\mathsf{u}}}_{jk}[n])$ is the \textit{generalized information density}, given by
\begin{align}
&\imath_s(q_{jk}[n],\by^{t,{\mathsf{u}}}_{jk}[n]) \nonumber \\
&\quad = -s \left|{\by^{t,{\mathsf{u}}}_{jk}[n] -\! \widehat{g}^{jk}_{jk} q_{jk}[n]}\right|^2 \!+\! \frac{s|\by^{t,{\mathsf{u}}}_{jk}[n]|^2}{1\!+\!s\rhou_{jk}|\widehat{g}^{jk}_{jk}|^2} \nonumber \\
&\quad\quad+ \log(1\!+\!s\rhou_{jk}|\widehat{g}^{jk}_{jk}|^2) \ ,
\label{eq:simple_infodens}
\end{align}
for all $s\!>\!0$. In~\eqref{eq:rcus_fading} the expectation is taken over the distribution of $g^{jk}_{jk}$, and the probability is computed with respect to the downlink data symbol $\{q_{jk}[n]\}_{n=1}^{\nd}$, the effective additive noise $\{z_{jk}[n]\}_{n=1}^{\nd}$, and the random variable $r$.
The evaluation of the upper bound in~\eqref{eq:rcus_fading} entails a very demanding numerical computation to firstly obtain the probability, and then to numerically tighten the upper bound value to the low error probability target of the URLLC use case by optimizing with respect to $s$. 

Luckily, we can reliably approximate the right-hand side of~\eqref{eq:rcus_fading} in closed form, hence with a significant relief of the computational burden, by using the \textit{saddlepoint} approximation provided in~\cite[Th. 2]{Ostman2021}.

The existence of a saddlepoint approximation is guaranteed by the fact that the third derivative of the \textit{moment-generating} function of $-\imath_s(q_{jk}[n],\by^{t,{\mathsf{u}}}_{jk}[n])$ exists in a neighborhood of zero delimited by the values $\underline{\varepsilon} \!<\! 0 \!<\! \overline{\varepsilon}$ given by~\cite[Appendix B]{Ostman2021}
\begin{align}
  \underline{\varepsilon} &= -\frac{\sqrt{(\zeta_{\mathsf{b}}-\zeta_{\mathsf{a}})^2 + 4\zeta_{\mathsf{a}}\zeta_{\mathsf{b}}(1-\mu)}+\zeta_{\mathsf{a}}-\zeta_{\mathsf{b}}}{2\zeta_{\mathsf{a}}\zeta_{\mathsf{b}}(1-\mu)}\label{eq:RoC_values_A} \ ,\\
  \overline{\varepsilon} &= \frac{\sqrt{(\zeta_{\mathsf{b}}-\zeta_{\mathsf{a}})^2 + 4\zeta_{\mathsf{a}}\zeta_{\mathsf{b}}(1-\mu)} -\zeta_{\mathsf{a}}+\zeta_{\mathsf{b}}}{2\zeta_{\mathsf{a}}\zeta_{\mathsf{b}}(1-\mu)}\label{eq:RoC_values_B} \ ,
\end{align}
where
\begin{align}
  \zeta_{\mathsf{a}} &= s(\rhou_{jk} |g^{jk}_{jk}-\widehat{g}^{jk}_{jk}|^2+\sigma^2)\label{eq:betaA} \ , \\
  \zeta_{\mathsf{b}} &= \frac{s}{1+s\rhou_{jk}|\widehat{g}^{jk}_{jk}|^2} (\rhou_{jk}|g^{jk}_{jk}|^2 + \sigma^2)\label{eq:betaB} \ , \\
  \mu &= \frac{s^2 \left|{\rhou_{jk}  |g^{jk}_{jk}|^2 + \sigma^2-{(g^{jk}_{jk})}^{\ast}\widehat{g}^{jk}_{jk}\rhou_{jk}}\right|^2}{\zeta_{\mathsf{a}} \zeta_{\mathsf{b}} (1+s \rhou_{jk} |\widehat{g}^{jk}_{jk}|^2)} \ .
\label{eq:corr_coeff_SISO}
\end{align}
The saddlepoint approximation hinges on the \textit{cumulant-generating} function of $-\imath_s(q_{jk}[n],\by^{t,{\mathsf{u}}}_{jk}[n])$ given by
\begin{align}
\upsilon(\varepsilon) = \log \EX{e^{-\varepsilon \imath_s(q_{jk}[n],\by^{t,{\mathsf{u}}}_{jk}[n])}} \ ,
\label{eq:cgf_def}
\end{align}
on its first derivative $\upsilon'(\zeta)$, and second derivative $\upsilon''(\zeta)$, for all $\varepsilon \in (\underline{\varepsilon},\overline{\varepsilon})$
\begin{align}
\upsilon(\varepsilon) &=\! -\varepsilon\log(1+s\rhou_{jk}|\widehat{g}^{jk}_{jk}|^2) \nonumber \\
&\quad- \log (1+(\zeta_{\mathsf{b}}-\zeta_{\mathsf{a}})\varepsilon-\zeta_{\mathsf{a}}\zeta_{\mathsf{b}}(1-\mu)\varepsilon^2)\label{eq:cgf}\\
\upsilon'(\varepsilon) &=\! -\log(1+s\rhou_{jk}|\widehat{g}^{jk}_{jk}|^2) \nonumber \\
&\quad- \frac{(\zeta_{\mathsf{b}}-\zeta_{\mathsf{a}}) -2\zeta_{\mathsf{a}}\zeta_{\mathsf{b}}(1-\mu)\varepsilon}{1+(\zeta_{\mathsf{b}}-\zeta_{\mathsf{a}})\varepsilon -\zeta_{\mathsf{a}}\zeta_{\mathsf{b}}(1-\mu)\varepsilon^2} \label{eq:cgf_1} \\
\upsilon''(\varepsilon) &=\! \left[\frac{(\zeta_{\mathsf{b}}-\zeta_{\mathsf{a}}) -2\zeta_{\mathsf{a}}\zeta_{\mathsf{b}}(1-\mu)\varepsilon}{1+(\zeta_{\mathsf{b}}-\zeta_{\mathsf{a}})\varepsilon-\zeta_{\mathsf{a}}\zeta_{\mathsf{b}}(1-\mu)\varepsilon^2}\right]^2 \nonumber \\
&\quad+ \frac{2\zeta_{\mathsf{a}}\zeta_{\mathsf{b}}(1-\mu)}{1+(\zeta_{\mathsf{b}}-\zeta_{\mathsf{a}})\varepsilon -\zeta_{\mathsf{a}}\zeta_{\mathsf{b}}(1-\mu)\varepsilon^2}. \label{eq:cgf_2}
\end{align} 
Let $m\!=\!e^{\nd R}$ for some strictly positive transmission rate $R\!=\!(\log m)/\nd$, and let $\varepsilon\in(\underline{\varepsilon},\overline{\varepsilon})$ be the solution to the equation $R\!=\!-\upsilon'(\varepsilon)$. Let  $I_{\mathsf{s}}$ be the  \textit{generalized mutual information}~\cite{Lapidoth2002} defined as $I_{\mathsf{s}}=\EX{\imath_s(q_{jk}[1],v_{jk}[1])}=-\upsilon'(0)$. Lastly, consider the \textit{critical rate}~\cite[Eq. (5.6.30)]{gallager1968} given by $R_{\mathsf{s}}^{\mathsf{cr}} = -\upsilon'(1)$.
Then, we have three possible saddlepoint approximations for the error probability upper bound~\cite{Ostman2021}. \\
If $\varepsilon\in[0,1]$, then $R_{\mathsf{s}}^{\mathsf{cr}} \leq R \leq I_{\mathsf{s}}$ and
\begin{align}
&\Prx{\sum_{n=1}^{\nd} {\imath_s(q_{jk}[n],\by^{t,{\mathsf{u}}}_{jk}[n])} \leq \log\frac{e^{\nd R}-1}{r}} \nonumber \\
&\quad\approx\! e^{\nd[\upsilon(\varepsilon)+\varepsilon R]}\left[\Qexp(\varepsilon)\!+\!\Qexp(1\!-\!\varepsilon)\right] \ , 
\label{eq:saddlepoint_U_pos}
\end{align}
where 
\begin{align}
\Qexp(\ell) \triangleq e^{\frac{1}{2}\nd\ell^2\upsilon''(\varepsilon)}Q\left(\ell\sqrt{\nd\upsilon''(\varepsilon)}\right) \ .
\label{eq:help_fcn_theta}
\end{align}
If $\varepsilon>1$, then $R < R_{\mathsf{s}}^{\mathsf{cr}}$ and
\begin{align}
&\Prx{\sum_{n=1}^{\nd} {\imath_s(q_{jk}[n],\by^{t,{\mathsf{u}}}_{jk}[n])} \leq \log\frac{e^{\nd R}-1}{r}} \nonumber \\
&\quad\approx e^{\nd[\upsilon(1)+ R]}\left[\widetilde{\Psi}_{\nd}(1,1)+\widetilde{\Psi}_{\nd}(0,-1)\right] \ , \label{eq:saddlepoint_U_pos_cr}
\end{align}
where 
\begin{align}
\widetilde{\Psi}_{\nd}(\ell_1,\ell_2) &\!\triangleq\! e^{\nd \ell_1\left[R_{\mathsf{s}}^{\mathsf{cr}}-R+\frac{1}{2}\upsilon''(1)\right]} \nonumber \\
&\quad\times\! Q\!\left(\ell_1\sqrt{\nd\upsilon''(1)}\!+\!\ell_2\frac{\nd(R_{\mathsf{s}}^{\mathsf{cr}}\!-\!R)}{\sqrt{\nd\upsilon''(1)}}\right).
\label{eq:help_fcn_theta_2}
\end{align}
If $\varepsilon<0$, then $R > I_{\mathsf{s}}$ and
\begin{align}
&\Prx{\sum_{n=1}^{\nd} {\imath_s(q_{jk}[n],\by^{t,{\mathsf{u}}}_{jk}[n])} \leq \log\frac{e^{\nd R}\!-\!1}{r}} \nonumber \\
&\quad\approx 1\!-\! e^{\nd[\upsilon(\varepsilon)+\varepsilon R]}\left[\Qexp(-\varepsilon) \!-\! \Qexp(1\!-\!\varepsilon)\right].\label{eq:saddlepoint_U_neg}
\end{align} 
The saddlepoint approximation is more accurate in the URLLC massive MIMO regime than the conventionally-used \textit{normal approximation}~\cite{Polyanskiy2010} as the former characterizes the exponential decay of the error probability, i.e., the error-exponent, as a function of the URLLC codeword length, and the transmission rate requirement $R$, while uses the \textit{Berry-Esseen central-limit theorem} (used in the normal approximation) to only characterize the multiplicative factor following the error-exponent term. The normal approximation, whose formulation directly involves the generalized mutual information, $I_{\mathrm{s}}$, but does not $R$, is accurate only when $I_{\mathrm{s}}$ is close to $R$. This operating regime does not hold for URLLC wherein $R$ is typically lower than $I_{\mathrm{s}}$ to accomplish the very low error probability targets. 
Once that the approximate upper bounds on the downlink error probability are obtained via saddlepoint approximation, we compute the \textit{downlink network availability}~\cite{Ostman2021}, $\eta^{\mathsf{dl}}$, as
\begin{align}\label{eq:network_avail}
\eta^{\mathsf{dl}} = \Prx{\epsilon^{\mathsf{dl}}_{jk} \leq \epsilon^{\mathsf{dl}}\sub{target}}
\end{align}
which measures the probability that the target error probability $\epsilon^{\mathsf{dl}}\sub{target}$ is satisfied by an arbitrary user $k$ in cell $j$, in presence of interfering users. While the expectation in the error probability definition is taken with respect to the small-scale fading and the effective additive noise, given a large-scale fading realization, the probability in the network availability definition is computed with respect to the large-scale fading (i.e., path loss, shadowing etc.). The expression of the network availability shown in~\eqref{eq:network_avail} holds for any choice of precoding scheme, any channel estimator and any channel distributions. Importantly, it accounts for any choice of coexistence technique between heterogeneous services, namely puncturing or superposition coding.

\section{PRECODING AND POWER CONTROL}
\label{sec:precoding-power-control}
The choice of the precoding scheme and of the downlink power allocation deeply affects the SE of the eMBB users and the network availability for the URLLC users. For the sake of comparison, we herein consider three precoding schemes and three power allocation strategies.
The general expression for the precoding vector intended for user $k$ in cell $j$ is given by 
\begin{align}
\bw_{jk} = \frac{\bv_{jk}}{\norm{\bv_{jk}}} \ ,
\end{align}
where the denominator serves to make the average power of the precoding vector unitary, and $\bv_{jk}$ is next characterized. \\
\textbf{Multi-cell MMSE (M-MMSE):}
\begin{align*}
\bv^{\mathsf{M-MMSE}}_{jk} \!=\!\! \left[\! \left(\sum\limits_{l=1}^L \widehat{\bH}^j_l \bP_l (\widehat{\bH}^j_l)\herm \!+\! \boldsymbol{\Upsilon}^j \!+\! \nvu\bI_M \!	\right)\inv\!\!\widehat{\bH}^j_j \bP_j \right]_{:,k} 
\end{align*}
where $\bP_l\!=\!\diag(p{li},\ldots,p_{lK}) \!\in\!\mR^{K\times K}$ is the matrix with the uplink transmit powers of all the users in cell $l$ as diagonal elements, $\boldsymbol{\Upsilon}^j = \sum\nolimits_{l=1}^L \sum\nolimits_{i=1}^K p_{li} \bC^j_{li}$, and $\widehat{\bH}^j_l \!=\! [ \hhat^j_{l1} \ldots \hhat^j_{lK}]$.
M-MMSE precoding provides a \textit{nearly optimal} downlink SE but requires each BS to acquire CSI and statistical CSI of all the users of the multi-cell system. Moreover, the computation of the precoding vector, which entails inverting a matrix $M\!\times\!M$, may be demanding for large BS arrays. Although impractical, M-MMSE precoding will serve as benchmark. \\     
\textbf{Regularized zero-forcing (RZF):}
\begin{align*}
\bv^{\mathsf{RZF}}_{jk} = \left[\widehat{\bH}^j_j \left((\widehat{\bH}^j_j)\herm \widehat{\bH}^j_j + \nvu \bP_j\inv \right)\inv \right]_{:,k} \ .  
\end{align*}
Compared to M-MMSE, RZF precoding requires each BS to estimate the channels of only its users. Moreover, computing the RZF precoding vector is computationally cheaper since the size of the matrix to be inverted is $K\!\times\!K$. However, RZF does only suppress the intra-cell interference while, unlike M-MMSE, does not provide to the users any protection mechanism against inter-cell interference and channel estimation error. \\
\textbf{Maximum Ratio (MR):} $\bv^{\mathsf{MR}}_{jk} = \hhat^j_{jk}.$ It is computationally the cheapest but performance-wise the worst precoding scheme. MR only aims at maximizing the power of the desired signal, providing no interference-suppression mechanism. MR will serve as lower bound on the performance.

Properly allocating the downlink power can make all the difference to meet the strict reliability requirements of the URLLC and to improve the SE of the eMBB users. Next, we provide three power allocation schemes that take into account the power budget at the BSs, the adopted eMBB-URLLC coexistence strategy and the URLLC activation pattern, which is known at the BS in the downlink operation. \\
\textbf{Equal power allocation (EPA):} It consists in setting
\begin{align}
\rhou_{ji} &= \rhoMax_j \frac{\Atji}{\Atj \Ke + \sum\limits_{k \in \setKu_j} \Atjk} \ ,~i\in\setKu_j \\
\rhoe_{jk} &= \rhoMax_j \frac{\Atj}{\Atj \Ke + \sum\limits_{i \in \setKu_j} \Atji} \ ,~k\in\setKe_j
\end{align}
to satisfy the per-BS power constraint in~\eqref{eq:data-power-constraint} with equality and allocate the same share of power to each user, regardless of its channel conditions and its service requirements. \\
\textbf{Weighted fractional power allocation (FPA):} it consists in setting the powers as
\begin{align}
\rhou_{ji} &=  \frac{\omega \rhoMax_j \Atji {(\beta^j_{ji})}^{\nu}}{(1\!-\!\omega)\Atj \!\sum\limits_{k \in \mathcal{K}^{\mathsf{e}}_j} {(\beta^j_{jk})}^{\nu} + \omega\!\!\sum\limits_{u \in \setKu_j} \!\Atju {(\beta^j_{ju})}^{\nu}} \ ,~i\in\setKu_j  \\[-.5ex]
\rhoe_{jk} &= \frac{(1-\omega) \rhoMax_j \Atj {(\beta^j_{jk})}^{\nu}}{(1\!-\!\omega)\Atj \sum\limits_{e \in \mathcal{K}^{\mathsf{e}}_j} {(\beta^j_{je})}^{\nu}  + \omega\!\! \sum\limits_{i \in \setKu_j} \Atji {(\beta^j_{ji})}^{\nu}} \ ,~k\in\setKe_j
\end{align}
where the weight $\omega \in (0,1)$ adjusts the amount of downlink power to be allocated to the URLLC users, while $\nu$ establishes the power control policy as a function of the average channel gain. An opportunistic power allocation is attained by setting $\nu\!>\!0$, with which more power is allocated to the users with better channel conditions. Conversely, fairness is supported by setting $\nu\!<\!0$, with which more power is allocated to the users with worse channel conditions.   
If $\omega\in (0.5,1)$ a larger share of power is allocated to the URLLC users rather than to the eMBB users, whereas it is the other way around if $\omega\in (0,0.5)$. Notice that, if $\nu\!=\!0$ and $\omega\!=\!0.5$, then the FPA reduces to the EPA. \\   
\textbf{Optimal power allocation (OPA) for max product SINR:}
The powers are the solution of the optimization problem
\begin{subequations} \label{prob:P1}
\begin{align}	
  \mathop {\text{maximize}}\limits_{\{\rhos_{jk}\}} & \quad \prod\limits^L_{j=1} \prod\limits^K_{k=1} \mathsf{SINR}^{t,\mathsf{s}}_{jk}  \label{prob:P1:obj} \\
  \text{s.t.} &\quad \sum\limits_{k = 1}^{K} \varrho^t_{jk} \leq \rhoMax_j \ , \forall j, \label{prob:P1:C1}
\end{align}
\end{subequations}%
where the superscript $\mathsf{s} \!=\! \mathsf{e}$ if user $k\in\setKe_j$, $\mathsf{s} \!=\! \mathsf{u}$ otherwise, and $\varrho^t_{jk}$ is given in~\eqref{eq:varrho-def}. Without further entangling the notation in~\eqref{prob:P1}, we remark that the SINR of inactive users is fictitiously set to 1 to preserve the optimization problem formulation. This power allocation strategy treats all the users as eMBB users, hence it would be optimal if there would be no URLLC users active in a given slot, by maximizing a lower bound on the sum SE of the multi-cell system. Although the SINR expression in~\eqref{eq:eMBB:SINR} is meaningless when applied to a URLLC user, we can still heuristically plug the URLLC powers resulting from~\eqref{prob:P1} into the error probability analysis and motivate this approach by looking at the performance. 
All the considered power allocation schemes, in principle, run on a slot-basis in order to adapt the power coefficients to the URLLC activation pattern. Fortunately, these schemes only rely on the knowledge of the statistical CSI which allows to pre-compute some power coefficients or to keep the power allocation for multiple slots/frames in case of no macroscopic changes in the propagation environment.
Unlike the EPA and the FPA schemes, the OPA scheme requires a certain degree of cooperation among the BSs which must send statistical CSI to let a central processing unit (e.g., a \textit{master} BS) compute the SINR of all the users and solve the optimization problem, and feed them back with the power coefficients to use. This would introduce intolerable delay for the URLLC users. 
Moreover, solving problem~\eqref{prob:P1}, although efficiently as a geometric program~\cite[Th. 7.2]{massivemimobook}, is unlikely to be doable within a time-slot, especially for crowded networks. 
Hence, the OPA scheme is of limited practical use, but will serve for benchmarking purposes.   

\section{SIMULATION RESULTS}
\label{sec:simulation-results}
In this section, we present and discuss the results of our simulations in which the coexistence of eMBB and URLLC is deeply analyzed under different setups. Specifically, we shed the light on the impact of different factors on the performance, such as the transmission technique and the precoding scheme, the power control strategy, the imperfect CSI and estimation overhead, the pilot contamination, the length and number of slots in a TDD frame, and the characteristics of the URLLC activation pattern. 

Our simulation scenario consists of a multi-cell massive MIMO system with $L\!=\!4$ cells. Each cell covers a nominal area of 500$\times$500 squared meters, and is served by a BS, placed at the cell center, equipped with a uniform linear array (ULA) with $M\!=\!100$ equispaced half-wavelength antenna elements. A wrap-around topology is implemented as in~\cite[Sec. 4.1.3]{massivemimobook}. 
The users are dropped uniformly at random over the coverage area but at a minimum distance of 25 m from the BS. In addition, we assume that the URLLC users are distributed uniformly at random in an area of 125$\times$125 squared meters that surrounds the BS.
A random realization of the user locations determines a set of large-scale fading coefficients and constitutes a snapshot of the network. For a given network snapshot the achievable downlink SEs of the active eMBB users are computed according to~\eqref{eq:eMBB:SE}, while the downlink error probabilities of the URLLC users are obtained according to the approximations~\eqref{eq:saddlepoint_U_pos}-\eqref{eq:saddlepoint_U_neg}. 
The cumulative distribution function (CDF) of the SE and the network availability are then drawn over many network snapshots.
The channel correlation matrices are generated according to the popular \textit{local scattering} spatial correlation model~\cite[Sec. 2.6]{massivemimobook}, and we assume that the scattering is only localized around the users and uniformly distributed at random with delay spread 25$^{\circ}$ degrees~\cite{Ostman2021}. The average channel gain is obtained according to the non-line-of-sight macro cell 3GPP model for 2 GHz carriers~\cite{LTE2017}, and given in dB by
\begin{align*}
\beta_k \! = \! -35.3-37.6\log_{10}\left(\frac{d_k}{1~\text{m}}\right) + F_k
\end{align*} 
for an arbitrary user $k$ placed at a distance $d_k$ from its BS, and where $F_k\!\sim\!\G{0}{\sigmash^2}$ models the log-normal shadowing as an \iid~random variable with standard deviation $\sigmash\!=\!4$ dB. 
The transmission bandwidth is 20 MHz, and the receiver noise power equals -94 dBm both for the uplink and the downlink. Moreover, we let $\rhoMax_j\!=\! 46$ dBm, $ j\!=\!1,\ldots,L$, and the uplink transmit power, both for pilot and payload data, be 23 dBm for all the users. 
We assume that the URLLC packet consists of $b\!=\!160$ bits, yielding a transmission rate $R\!=\!b/\nd$, which is suitable for factory automation use cases, such as motion controls, and in line with the low latency requirements\cite[Annex A]{URLLCspecs3GPP}. Lastly, without loss of generality, we set $\tu\!=\!0$ as we only focus on the downlink performance. Unless otherwise stated, we consider TDD frames with length $\tc = 580$ channel uses, given by $\Tc\!=\!2$ ms and $\Bc\!=\!290$ kHz, which supports user mobility  up to 67.50 km/h. 

\begin{figure}
\centerline{\includegraphics[width=\columnwidth]{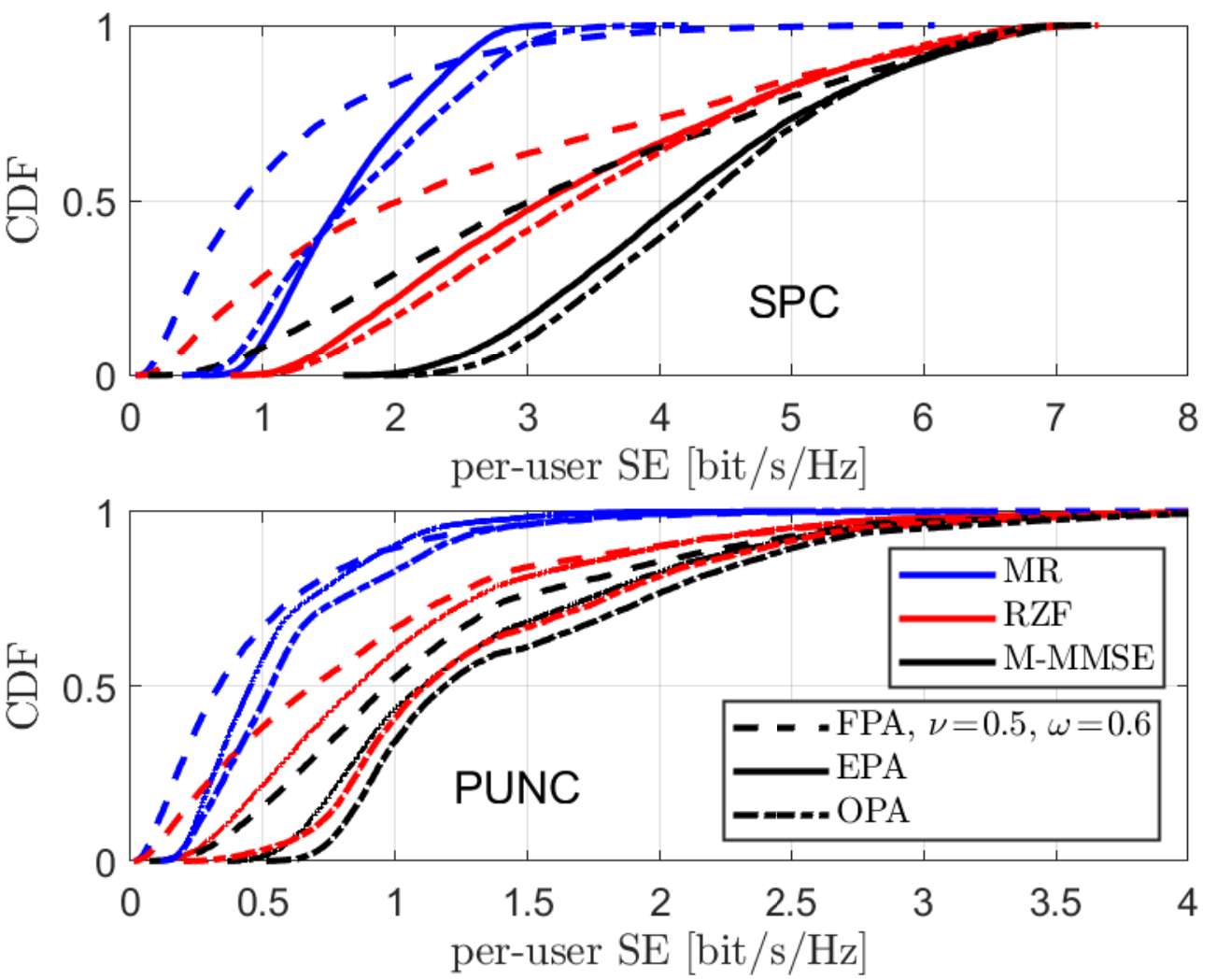}}
\caption{\label{fig:fig1a} CDFs of the achievable downlink SE per active eMBB user, for different transmission, precoding and power allocation strategies. Settings: $K\!=\!20$, $\alpha\!=\!0.2$, $\au\!=\!10^{-0.5}$, $\tp\!=\!80$, $T\!=\!5$, $\nd\!=\!100$.}
\end{figure}

In the first set of simulations we consider the following setup: $K\!=\!20$, $\alpha\!=\!0.2$, $\au\!=\!10^{-0.5}$, $\tp\!=\!80$ (no pilot contamination), $T\!=\!5$ slots of length $\nd\!=\!100$ channel uses. In~\Figref{fig:fig1a} we plot the CDFs of the achievable downlink SE per ``active'' eMBB user obtained for different precoding and power allocation strategies, both for superposition coding (top subfigure) and puncturing technique (bottom subfigure).        
Under these assumptions, SPC is greatly superior than PUNC, precoding and power allocation strategies being equal. M-MMSE with OPA gives, as expected, the best SE but EPA performs almost equally well, regardless of the precoding scheme. RZF provides a practical excellent trad-off between M-MMSE and MR.  
These results suggest that we are approximately operating in an interference-free scenario, thanks to the full and partial interference-suppression mechanism provided by M-MMSE and RZF, respectively. As per the FPA strategy, in these simulations we have selected $\nu\!=\!0.5$ to promote an opportunistic power allocation and $\omega\!=\!0.6$ to prioritize the URLLC users. Such a choice does not favor the eMBB users and justify the worst performance of FPA among the considered strategies when SPC is applied. 

\begin{figure}
\centerline{\includegraphics[width=\columnwidth]{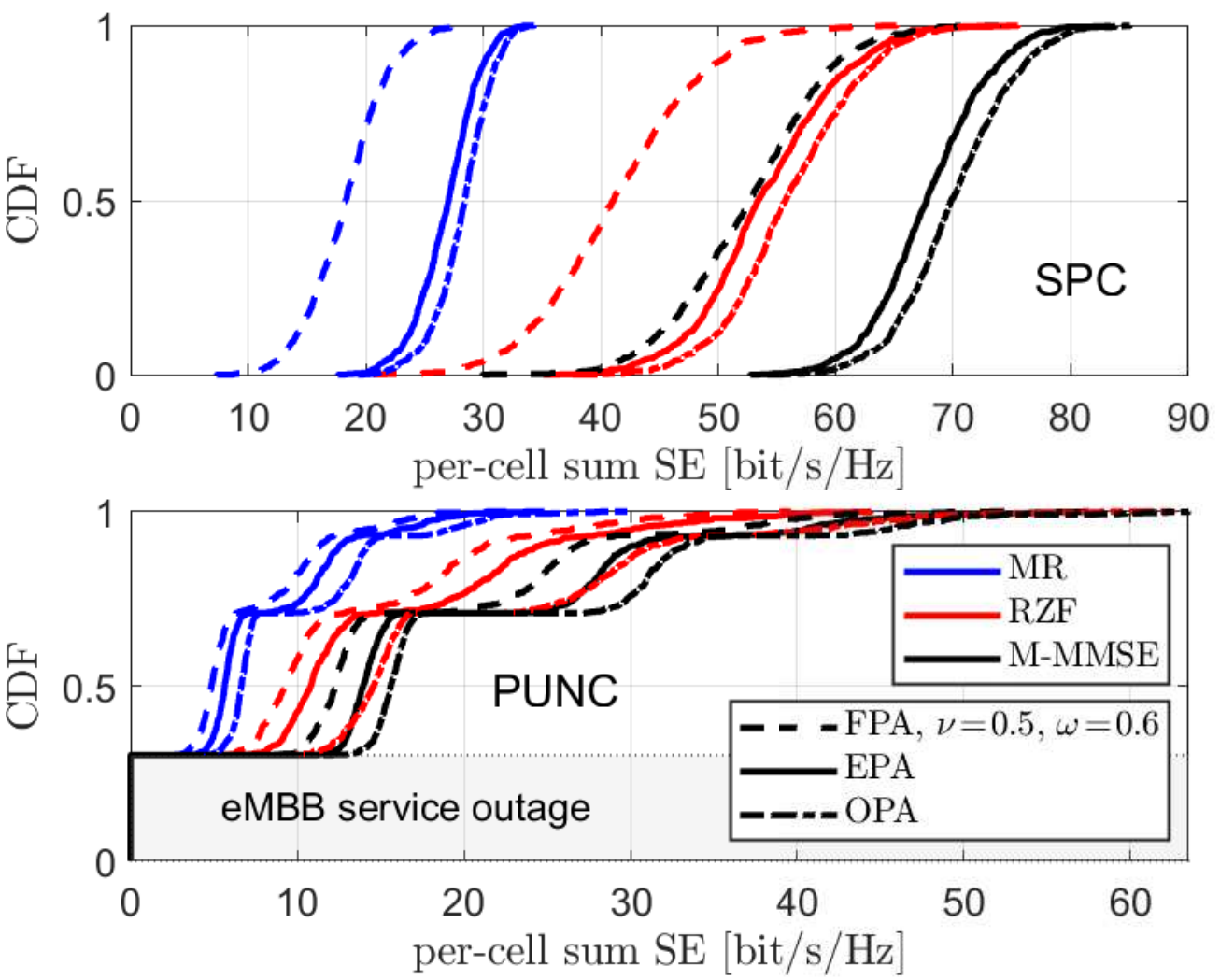}}
\caption{\label{fig:fig1a_bis} CDFs of the achievable downlink sum SE per cell, for different transmission, precoding and power allocation strategies. Settings: $K\!=\!20$, $\alpha\!=\!0.2$, $\au\!=\!10^{-0.5}$, $\tp\!=\!80$, $T\!=\!5$, $\nd\!=\!100$.}
\end{figure}

Same conclusions hold for the results shown in~\Figref{fig:fig1a_bis} where the CDFs of the corresponding sum SE per cell are illustrated. 
In these figures, we mainly emphasize the eMBB \textit{service outage} likely occurring when PUNC is adopted. We define the eMBB service outage, under PUNC operation, as
\begin{align*}
\varsigma_{\, \mathrm{out}} = \Prx{\sum_{k \in \setKe_j} \mathsf{SE}^{\mathsf{e}}_{jk} = 0},\quad j\!=\!1,\ldots,L \ , 
\end{align*} 
where the probability is computed with respect to the large-scale fading. This probability for a BS to provide no service in a TDD frame to its eMBB users depends on the activation pattern of the URLLC users and the number of slots per frame. We will discuss this aspect in detail later. Under the settings considered in~\Figref{fig:fig1a_bis}, the eMBB service outage is quite significant as amounts to about 30\%. 

\begin{figure}
\centerline{\includegraphics[width=\columnwidth]{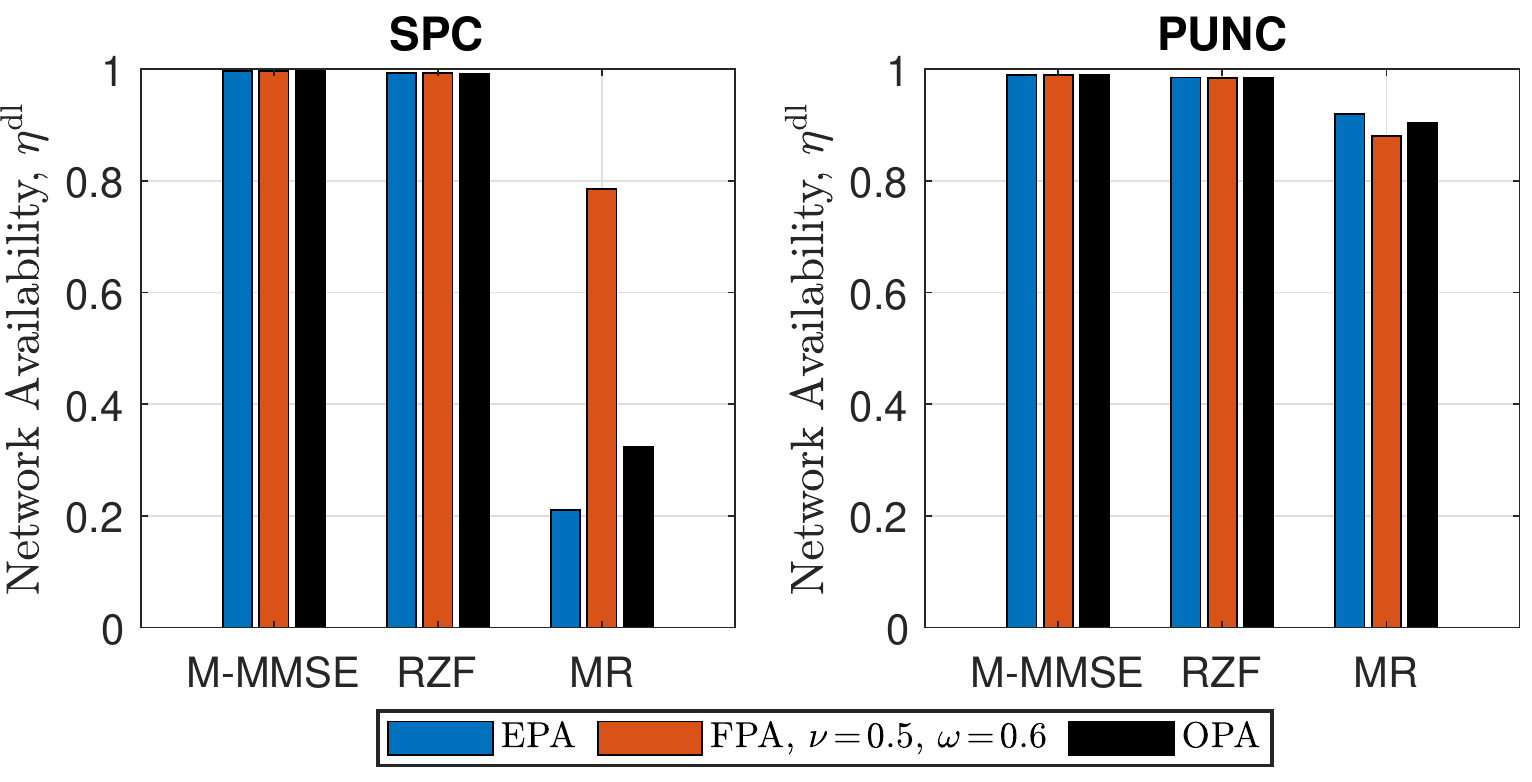}}
\caption{\label{fig:fig1_b} Network availability for different transmission, precoding and power allocation strategies. Settings: $K\!=\!20$, $\alpha\!=\!0.2$, $\au\!=\!10^{-0.5}$, $\tp\!=\!80$, $T\!=\!5$, $\nd\!=\!100$. }
\end{figure}

\begin{figure}
\centerline{\includegraphics[width=\columnwidth]{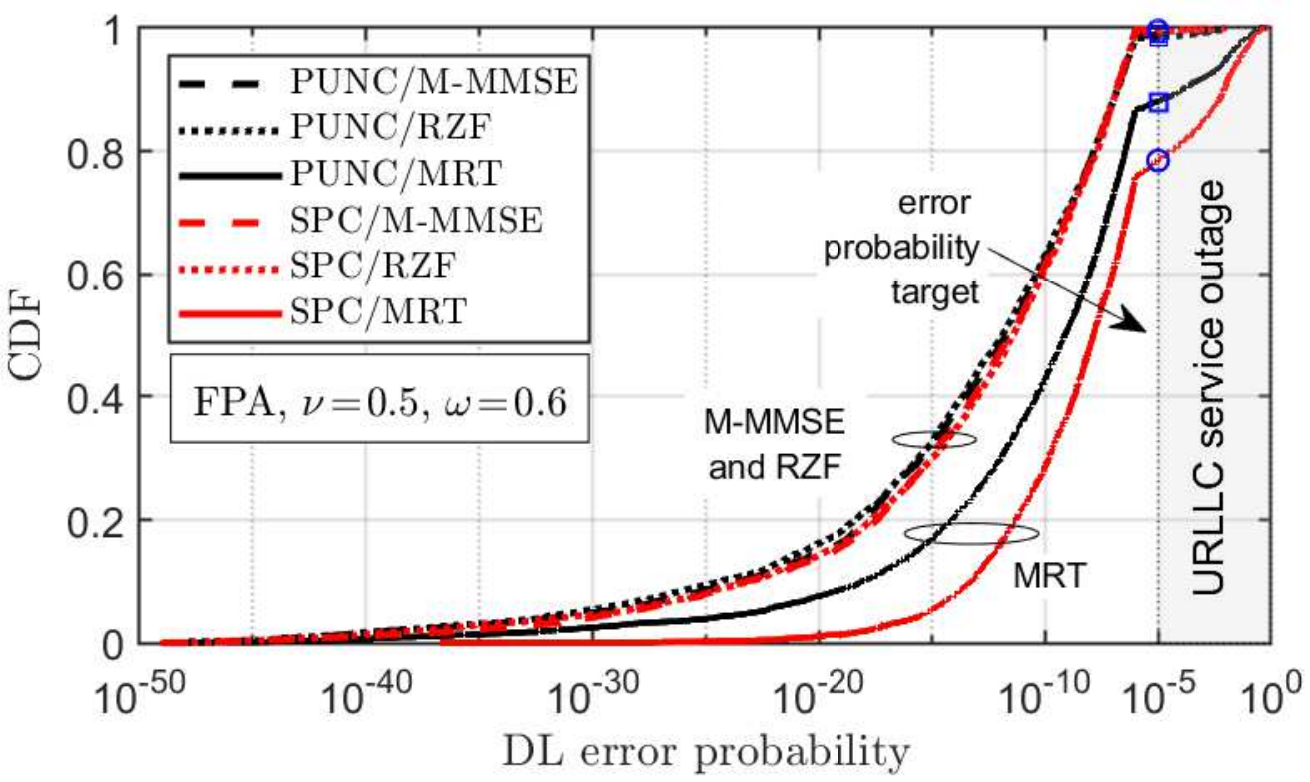}}
\caption{\label{fig:fig1_c} Downlink per-user error probability for different transmission and precoding strategies. Settings: EPA, $K\!=\!20$, $\alpha\!=\!0.2$, $\au\!=\!10^{-0.5}$, $\tp\!=\!80$, $T\!=\!5$, $\nd\!=\!100$.}
\end{figure}

In~\Figref{fig:fig1_b} we move to the URLLC performance by showing the downlink network availability achieved when $\epsilon^{\mathsf{dl}}\sub{target}\!=\!10^{-5}$. 
Despite the interference caused by the eMBB users when SPC is performed, both M-MMSE and RZF are able to provide levels of network availability close to one, in line with PUNC, revealing a great ability of suppressing the interference and supporting high reliability. Conversely, MR provides poor performance in SPC when EPA or OPA (which is optimal for the eMBB users) schemes are used. Notice that, our choice for the parameters of the FPA scheme pays off for the combination SPC/MR. The network availability values shown in~\Figref{fig:fig1_b} are obtained by the error probabilities whose CDFs are illustrated in~\Figref{fig:fig1_c}.   
To better understand its meaning, the network availability is given by the cross-point between the CDF of the per-user error probability and the vertical line representing the error probability target value, as~\Figref{fig:fig1_c} highlights (blue circle markers). From this set of simulations, we conclude that SPC is clearly superior to PUNC in terms of SE yet providing very high network availability, when M-MMSE or RZF are carried out. If MR is the only viable option (for instance due to strict complexity or hardware constraints), then SPC with FPA, upon properly setting the design parameters $\nu$ and $\omega$,  is an effective choice to keep the network availability high while preventing any eMBB service outage. 

\begin{figure}
\centerline{\includegraphics[width=\columnwidth]{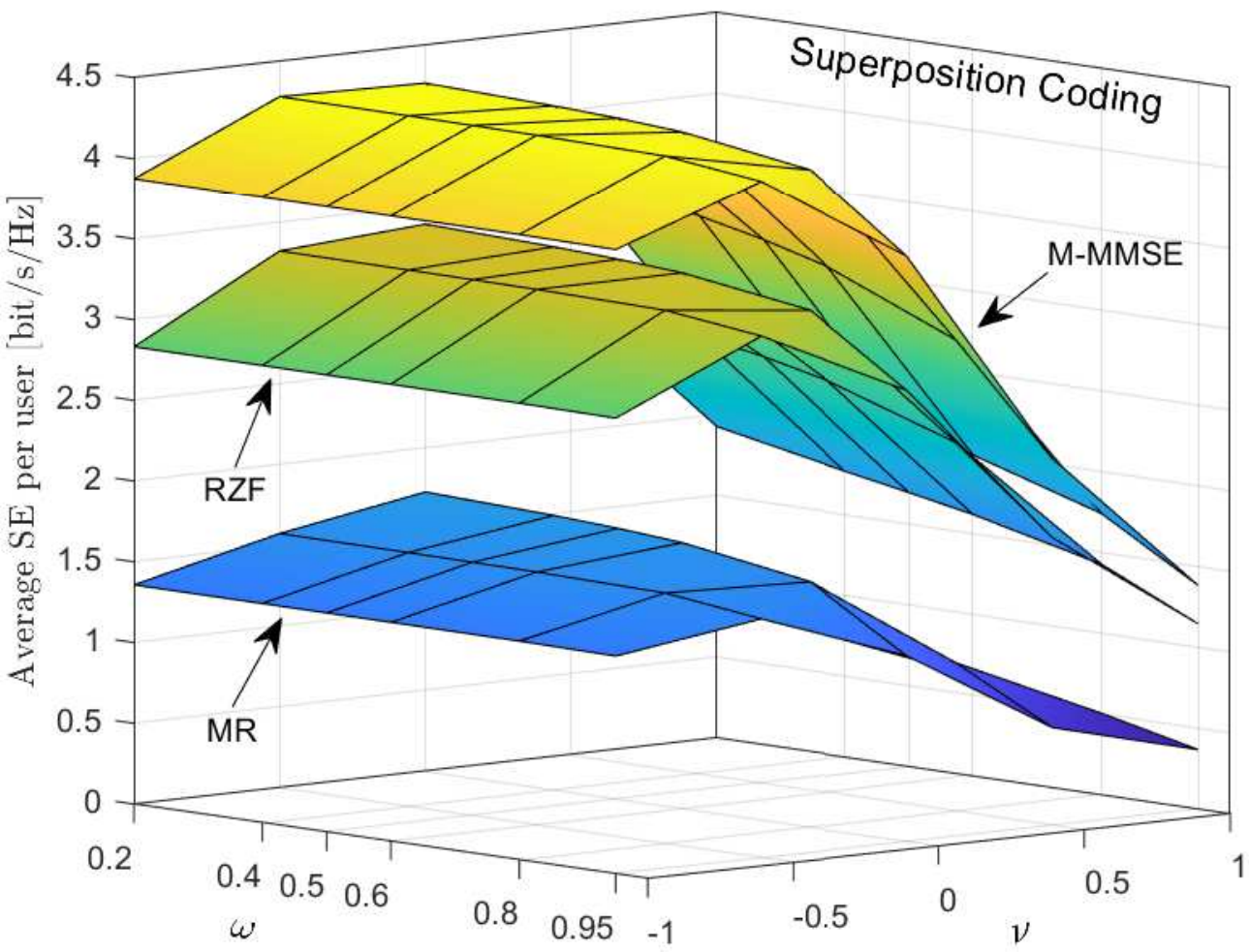}}
\caption{\label{fig:Fig2_a} Average per-user SE achieved by SPC with FPA, for different precoding schemes and values of $\nu$, $\omega$. The average is taken over 200 network snapshots. Settings: $K\!=\!20$, $\alpha\!=\!0.2$, $\au\!=\!10^{-0.5}$, $\tp\!=\!80$, $T\!=\!5$, $\nd\!=\!100$. }
\end{figure}

In this regard, we now focus on how to select $\nu$ and $\omega$ appropriately. By using the same settings as in the first set of simulations, in~\Figref{fig:Fig2_a} we plot the average per-user SE assuming SPC and different precoding schemes with FPA as $\nu$ and $\omega$ vary.
From the eMBB user perspective, it is preferable setting a small value for $\omega$, and $\nu$ in the interval $[-0.5,0]$. While the former is trivial, the latter needs further discussions. Indeed, recall that positive values for $\nu$ enable allocating more power to users with better channel conditions. Since we assume the URLLC users are uniformly distributed in a smaller area surrounding the BSs, it is very likely that they are closer to the BS than most of the eMBB users. Therefore, negative values for $\nu$ increase the fairness and improve eMBB users performance. Large values for both $\omega$ and $\nu$ excessively unbalance the power distribution in favor of the URLLC users, degrading the SE of the eMBB users. 

\begin{figure}
\centerline{\includegraphics[width=\columnwidth]{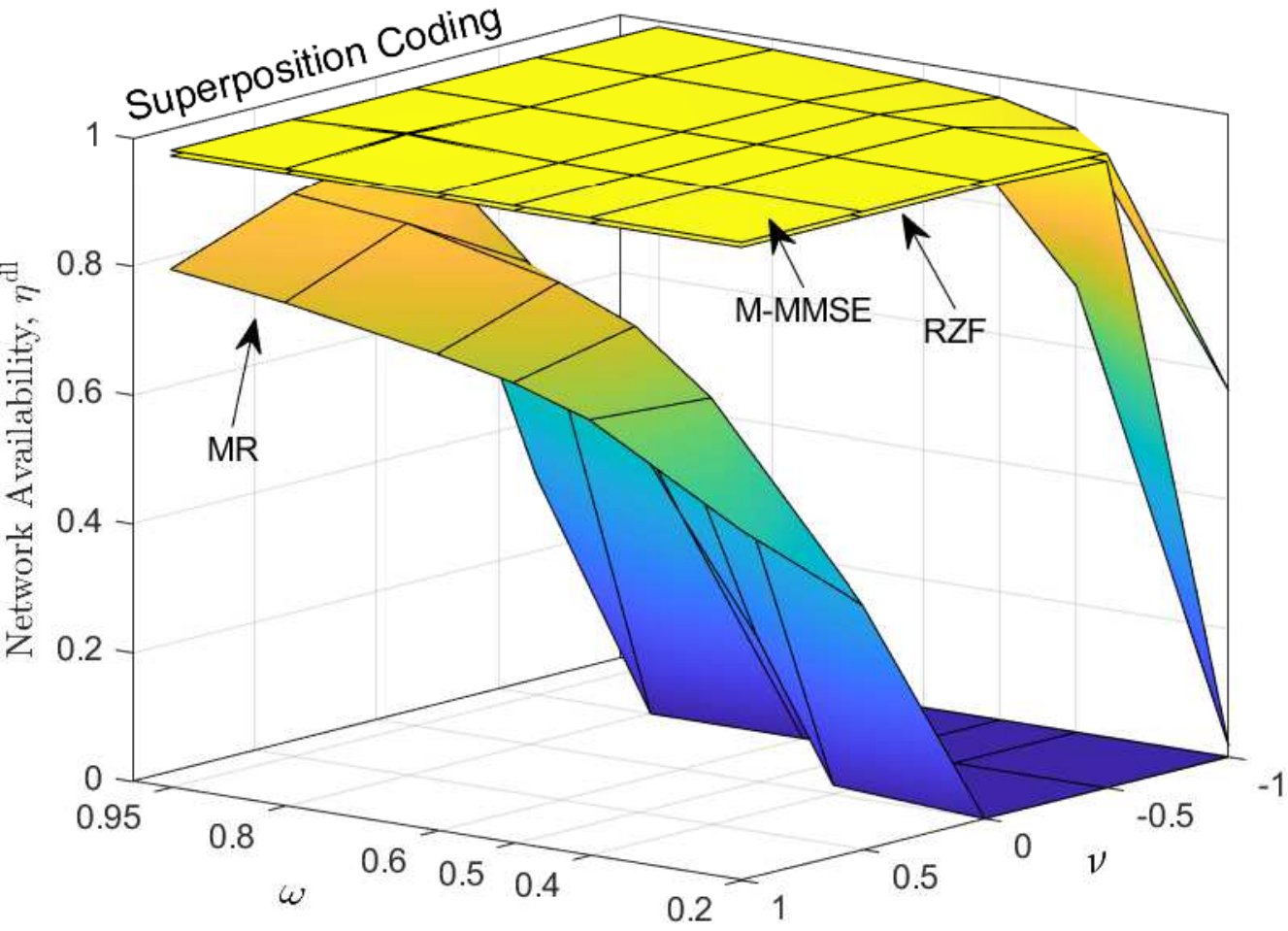}}
\caption{\label{fig:Fig2_b} Network availability achieved by SPC with FPA, for different precoding schemes and values of $\nu$, $\omega$. Settings: $K\!=\!20$, $\alpha\!=\!0.2$, $\au\!=\!10^{-0.5}$, $\tp\!=\!80$, $T\!=\!5$, $\nd\!=\!100$. }
\end{figure}

Conversely, small values for both $\omega$ and $\nu$ break down the network availability of the URLLC users in SPC operation, as clearly seen in~\Figref{fig:Fig2_b}.
Nevertheless, both M-MMSE and RZF are able to provide levels of network availability close to 1 except when $\nu\!=\!-1$, while MR is quite sensitive to this parameters tuning. Suppressing the multi-user interference is of a vital importance when SPC is adopted, and RZF, although not dealing with the inter-cell interference, is an excellent trade-off between performance and practicality. Fine-tuning the parameters of the FPA scheme yields satisfying performance when using MR. FPA becomes a valid, heuristic alternative to combat the multi-user interference whenever the latter cannot be removed by the precoding technique.

\begin{figure}
\centerline{\includegraphics[width=\columnwidth]{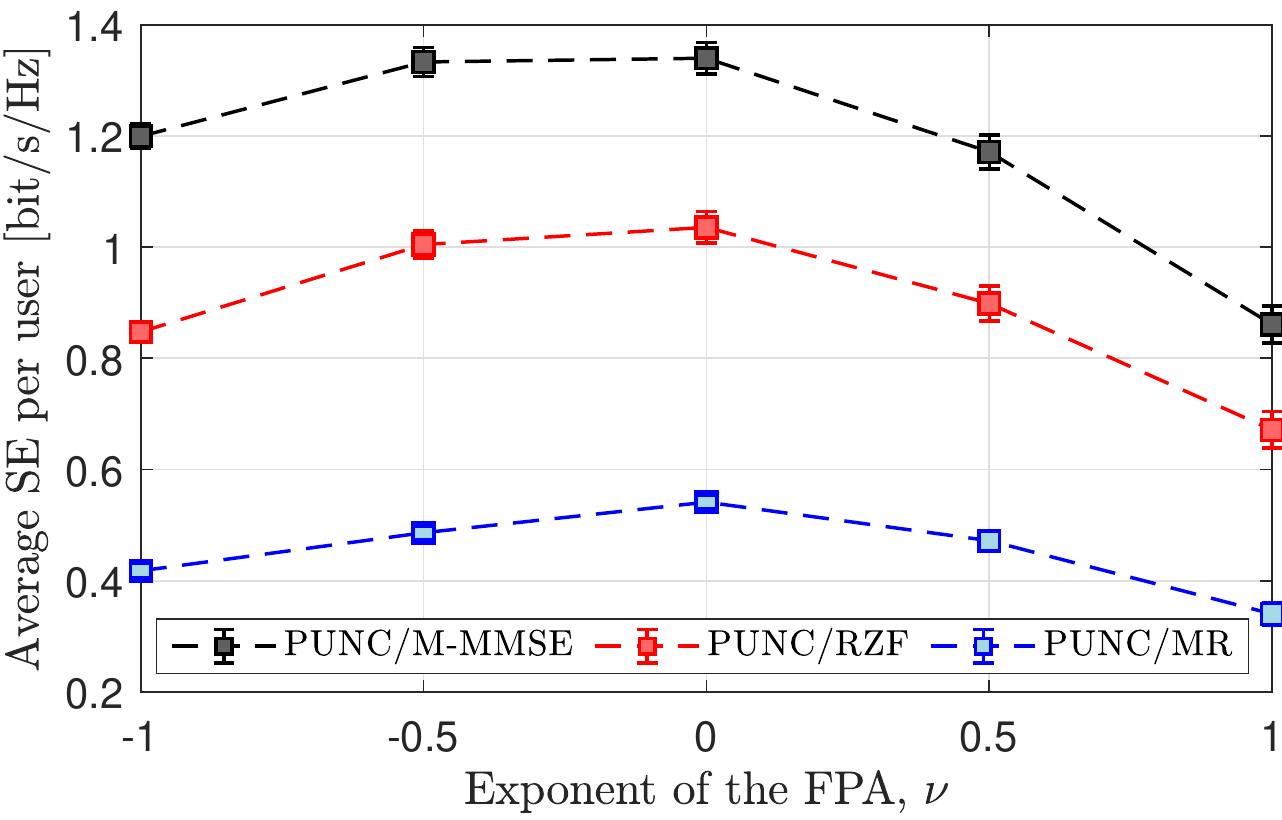}}
\caption{\label{fig:Fig2_c} Average per-user SE (with 95\% confidence interval) achieved by PUNC with FPA, for different precoding schemes and values of $\nu$. The average is taken over 200 network snapshots. Settings: $K\!=\!20$, $\alpha\!=\!0.2$, $\au\!=\!10^{-0.5}$, $\tp\!=\!80$, $T\!=\!5$, $\nd\!=\!100$. }
\end{figure}

\begin{figure}
\centerline{\includegraphics[width=\columnwidth]{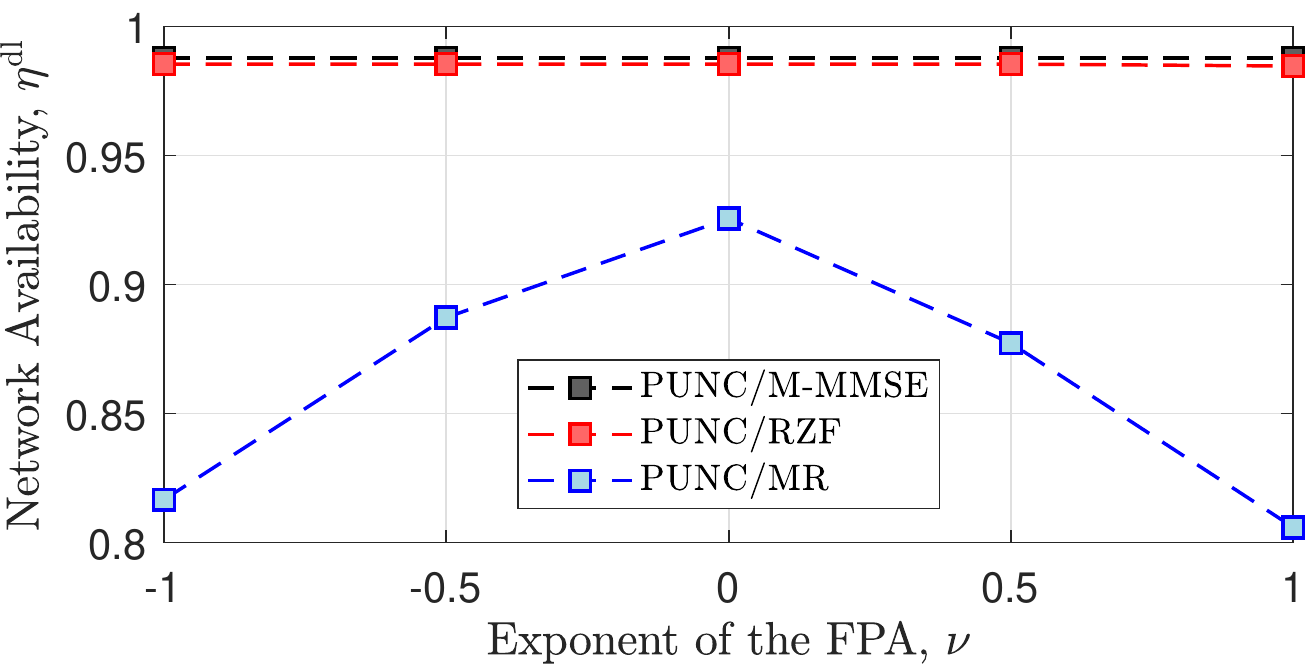}}
\caption{\label{fig:Fig2_d} Network availability achieved by PUNC with FPA, for different precoding schemes and values of $\nu$. Settings: $K\!=\!20$, $\alpha\!=\!0.2$, $\au\!=\!10^{-0.5}$, $\tp\!=\!80$, $T\!=\!5$, $\nd\!=\!100$. }
\end{figure}

Setting $\omega$ becomes pointless when using PUNC with FPA as only URLLC transmissions take place in the considered slot. Hence, in~\Figref{fig:Fig2_c} and~\Figref{fig:Fig2_d} we focus on the average SE per user and the network availability as only $\nu$ varies. 
For both cases we notice that an equal power allocation, i.e., $\nu\!=\!0$, is desirable. As per the SE of the eMBB users, negative values of $\nu$ support lower SEs (e.g., the 95\%-likely SE per user), hence the fairness among the users, while large positive values of $\nu$ support the peak SE in a greedy fashion, neglecting lower SEs. Therefore, $\nu\!=\!0$ is sound if the average SE is targeted, especially when the multi-user interference is partially or fully canceled.  
As per the network availability of the URLLC users, any choice of $\nu\in[-1,1]$ is solid as long as M-MMSE or RZF are employed, while the performance of MR is relatively penalized whenever a non-neutral choice for $\nu$ is taken. Presumably, the number of URLLC users simultaneously active in the same slot (resulting from the chosen values of $\alpha$ and $\au$) is such that the multi-user interference is not significant.

\begin{figure}
\centerline{\includegraphics[width=\columnwidth]{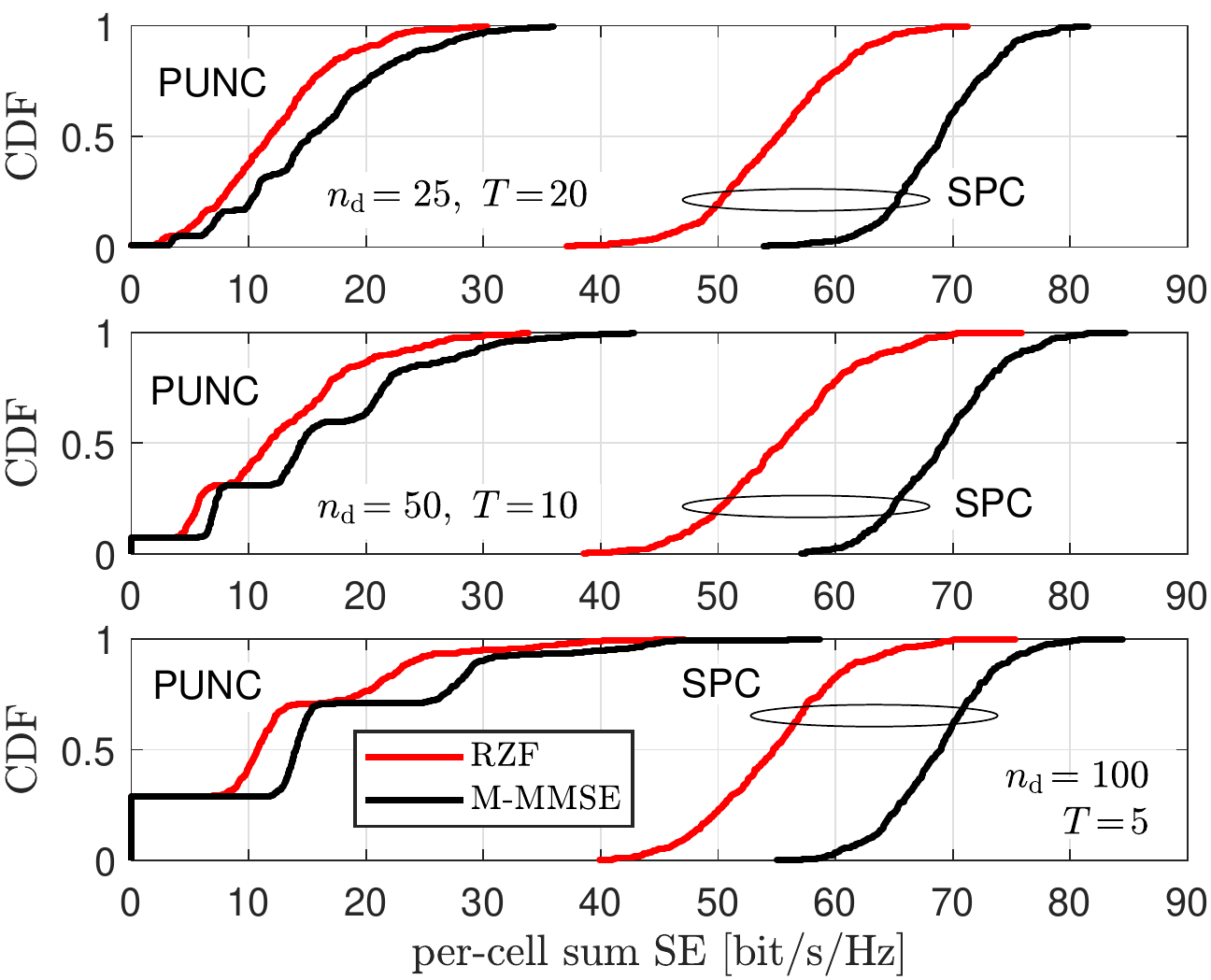}}
\caption{\label{fig:Fig3_a} CDFs of the achievable downlink sum SE per cell, for different transmission and precoding strategies, as the number of slots per frame varies. Settings: FPA with $\nu\!=\!0$ and $\omega\!=\!0.2$, $K\!=\!20$, $\alpha\!=\!0.2$, $\au\!=\!10^{-0.5}$, $\tp\!=\!80$.}
\end{figure}

Next, we evaluate the performance as a function of the number of the slots in a TDD frame, $T$, and the size of the slot, $\nd$, which in turn determines the URLLC codeword length. In this set of simulations and hereafter, we omit the results achieved by MR and only consider FPA with $\nu\!=\!0$ and $\omega\!=\!\alpha$ motivated by the previous results.
\Figref{fig:Fig3_a} shows the CDFs of the sum SE per cell, for three different setups: $(i)$ $\nd\!=\!25$, $T\!=\!20$, $(ii)$ $\nd\!=\!50$, $T\!=\!10$, and $(iii)$ $\nd\!=\!100$, $T\!=\!5$. 
The structure of the TDD frame has not a significant impact on the SE of the eMBB users when SPC is used. Conversely, that  deeply affects the per-cell sum SE in case of PUNC. Indeed, increasing the number of slots per frame makes the probability of having eMBB service outage smaller as it increases the opportunities for an eMBB user to find slots with no active URLLC users. This argument is supported by the results in~\Figref{fig:Fig3_a} in which the eMBB service outage equals 0.01, 0.0725 and 0.2875 when $T\!=20$, $T\!=\!10$ and $T\!=\!5$, respectively. On the other hand, with fewer slots, eMBB users might be active for longer time, thereby experiencing higher SE. This explains the larger variations of the per-cell sum SE as $T$ is decreased. 

\begin{figure}
\centerline{\includegraphics[width=\columnwidth]{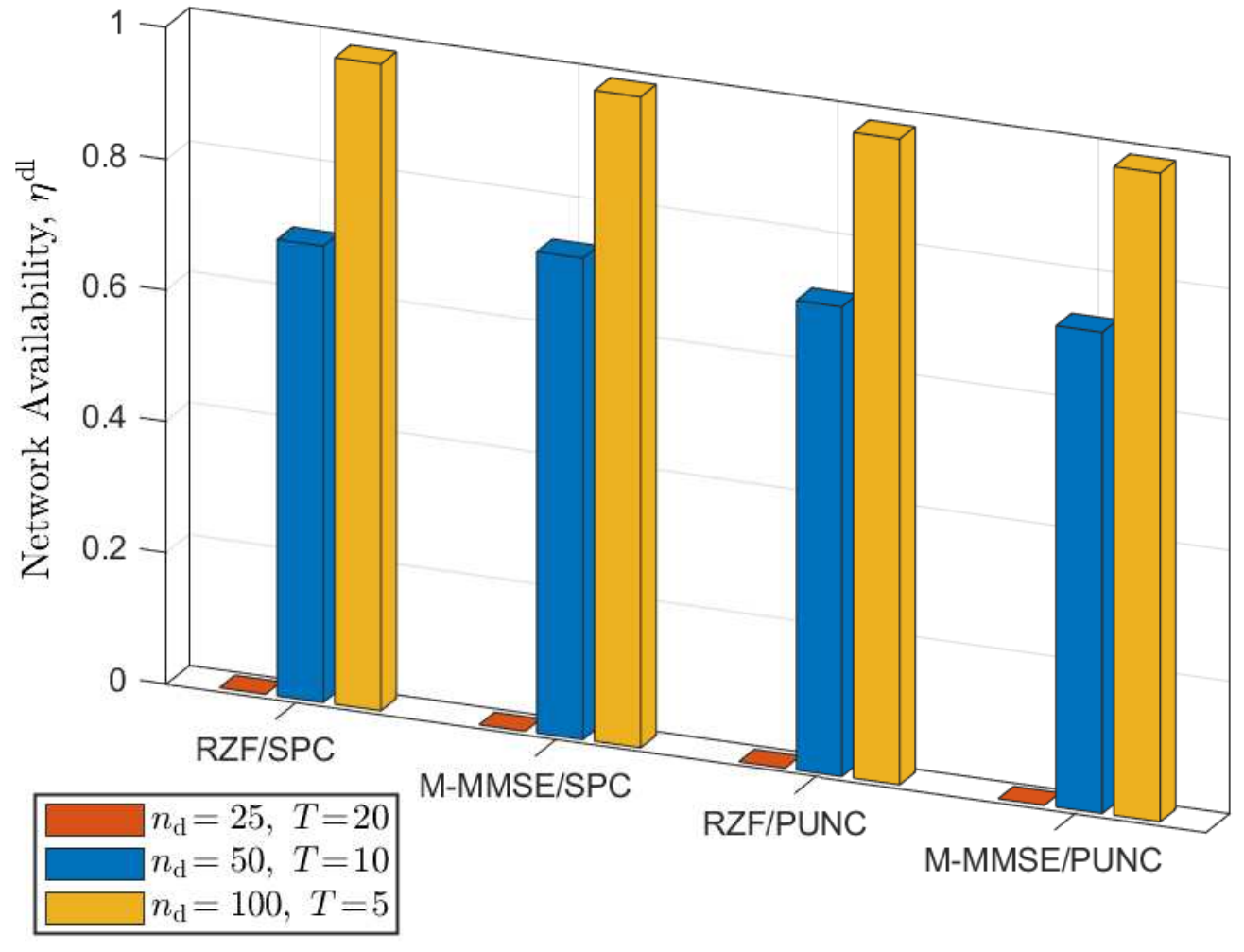}}
\caption{\label{fig:Fig3_b} Network availability, for different transmission and precoding strategies, as the length of the slot varies. Settings: FPA with $\nu\!=\!0$ and $\omega\!=\!0.2$, $K\!=\!20$, $\alpha\!=\!0.2$, $\au\!=\!10^{-0.5}$, $\tp\!=\!80$.}
\end{figure}

The length of the slot directly affects the performance of the URLLC users. As we can see in~\Figref{fig:Fig3_b}, the network availability increases drastically with the length of the slot (i.e., the URLLC codeword length).    
In fact, the length of the URLLC codeword determines the transmission rate of the URLLC users as $R\!=\!b/\nd$, thus the shorter the codeword the higher the rate requirement to be reliably achieved and, in turn, the larger the error probability.\footnote{The random-coding union bound in~\eqref{eq:rcus_fading} defines the error probability as the probability that the average generalized information density is smaller than the transmission rate requirement.} Again, SPC is the technique that overall guarantees the best performance to both the eMBB and URLLC users as its main limitation, namely the caused multi-user interference, is overcome by using interference-suppression-based precoding schemes. Lastly, although letting the URLLC transmissions span many channel uses is beneficial in terms of network availability, the latency requirements impose to localize the transmissions in time.

\begin{figure}
\centerline{\includegraphics[width=\columnwidth]{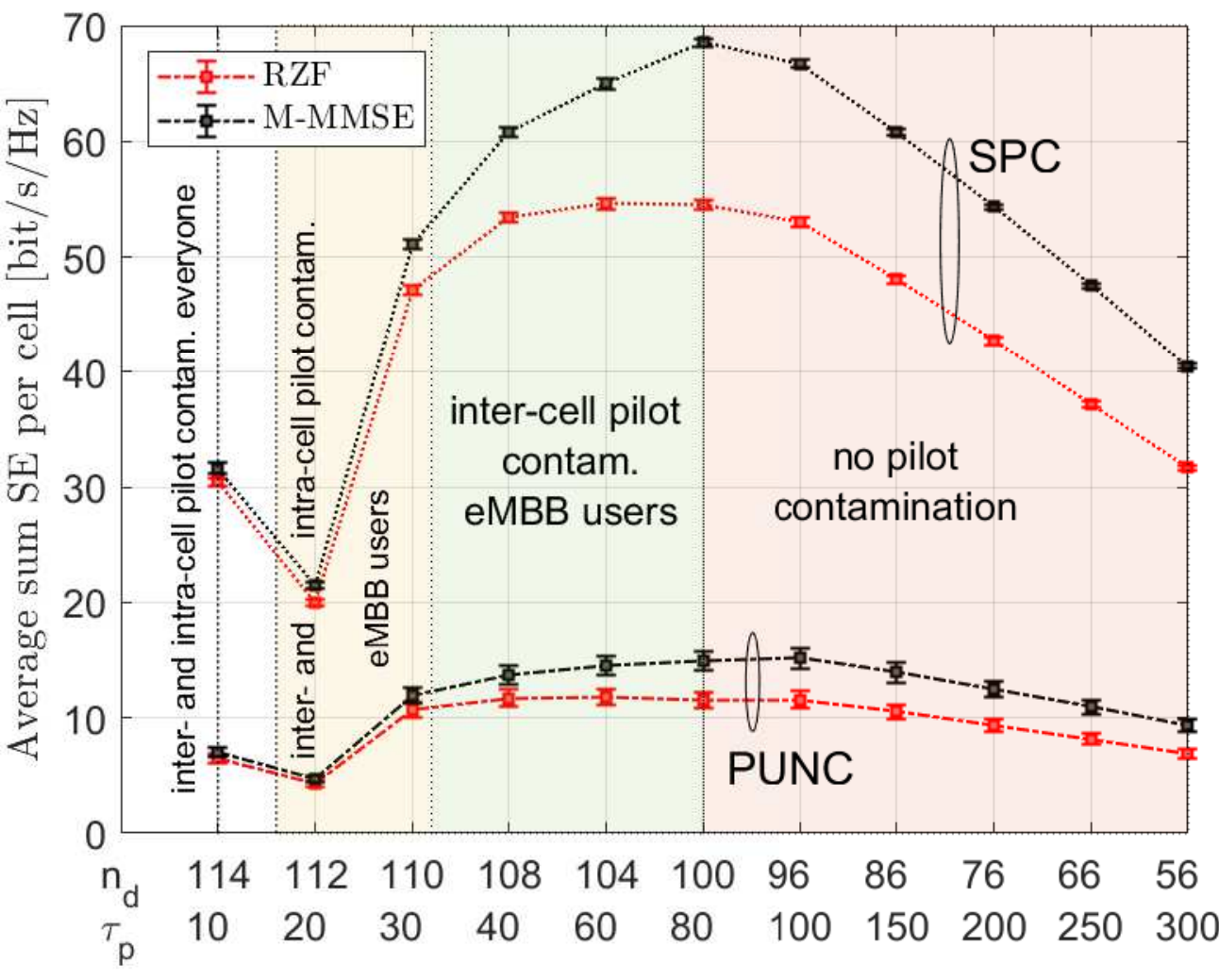}}
\caption{\label{fig:Fig4_a} Average SE per cell (with 95\% confidence interval), for different transmission and precoding strategies, as $\tp$ (and $\nd$) varies. The average is taken over 200 network snapshots. Settings: FPA with $\nu\!=\!0$ and $\omega\!=\!0.2$, $K\!=\!20$, $\alpha\!=\!0.2$, $\au\!=\!10^{-0.5}$, $\tc\!=\!580$, $T\!=\!5$. }
\end{figure}

Now, we move our focus on the impact of the pilot contamination and estimation overhead on the performance. By fixing the TDD frame length and the number of slots per frame, we vary the length of the uplink training, hence the number of available orthogonal pilots, and the length of each slot accordingly. In~\Figref{fig:Fig4_a} we show how the average sum SE per cell evolves in different operating regimes with respect to the uplink training length. 
In these simulations, we assume $K\!=\!20$, $\alpha\!=\!0.2$, $\tc\!=\!580$ and $T\!=\!5$. Small values of $\tp$ entails low channel estimation overhead but high levels of pilot contamination which reduces the effectiveness of the precoding. Our pilot assignment scheme preserves the performance of the URLLC users by assigning them unique pilots if available, otherwise pilots are assigned randomly and contamination hits any user indiscriminately. The maximum number of URLLC users potentially active in this scenario is, according to the chosen parameter, 16. Hence, pilots are assigned randomly when $\tp\!=\!10$ causing both intra- and inter-cell pilot contamination and providing a low sum SE per cell, namely about 30 bit/s/Hz with SPC and less than 10 bit/s/Hz with PUNC. The performance worsens when $\tp\!=\!20$ as the eMBB users have to share only 4 orthogonal pilots since the protection mechanism of the URLLC users is now triggered. As we increase the value of $\tp$, the intra-cell pilot contamination is primarily reduced by assigning orthogonal pilots to eMBB users of the same cell. If $\tp\!\geq\!32$ then intra-cell pilot contamination is prevented and the inter-cell interference among the eMBB users remains the only impairment. The sum SE per cell keep growing up to $\tp\!=\!80$, when all the users in the network are assigned mutual orthogonal pilots and the benefits of having no pilot contamination at all overcome the penalty from increasing the estimation overhead.  Trivially, there are no benefits in the channel estimation when further increasing $\tp$, while the estimation overhead turns to be expensive and drastically lowers the sum SE per cell. Finally, notice that RZF and M-MMSE provide essentially the same performance when both the intra- and inter-cell pilot contamination occur, because the ability of suppressing the multi-user interference is poor for both the schemes. 

\begin{figure}
\centerline{\includegraphics[width=\columnwidth]{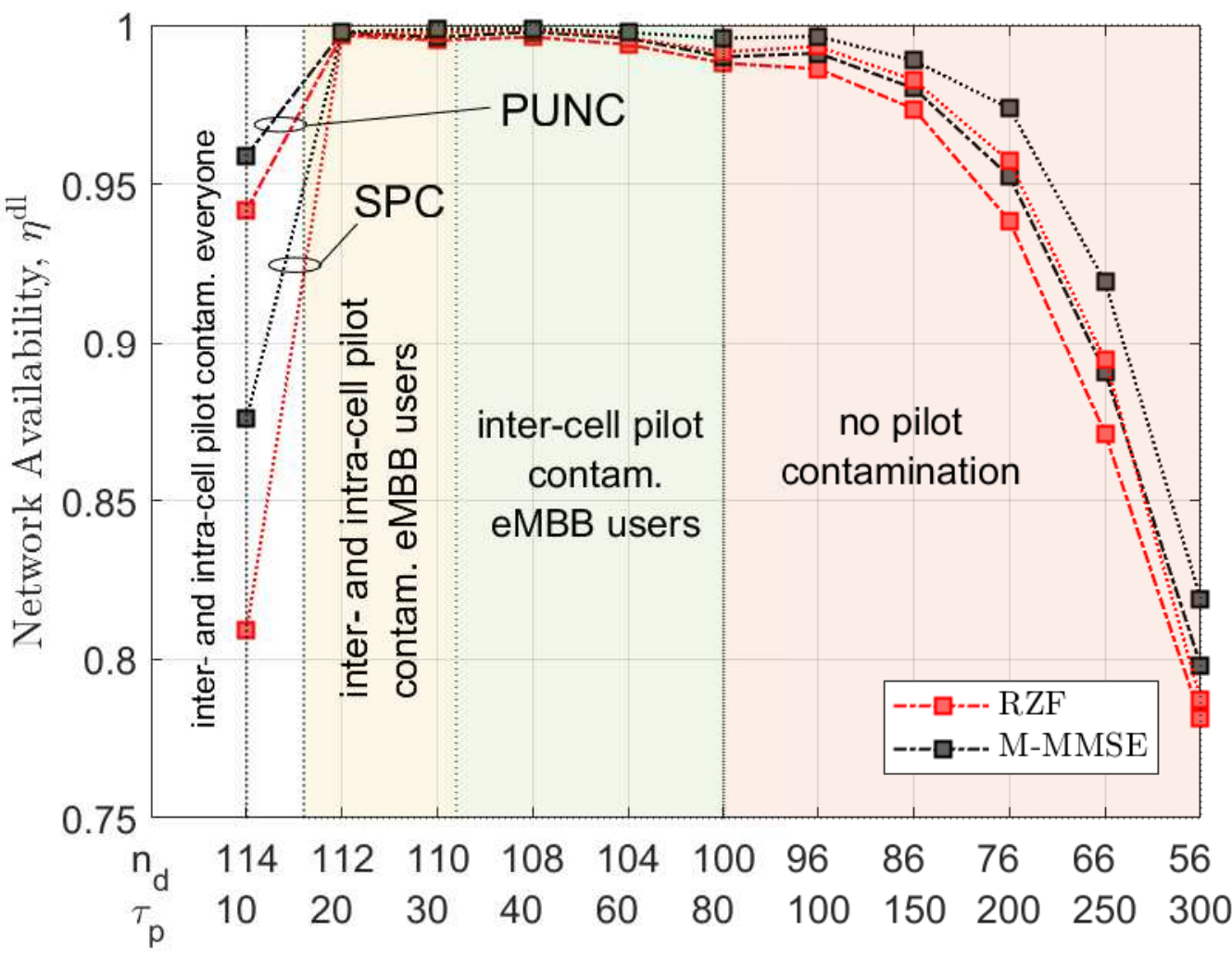}}
\caption{\label{fig:Fig4_b} Network availability, for different transmission and precoding strategies, as $\tp$ (and $\nd$) varies. Settings: FPA with $\nu\!=\!0$ and $\omega\!=\!0.2$, $K\!=\!20$, $\alpha\!=\!0.2$, $\au\!=\!10^{-0.5}$, $\tc\!=\!580$, $T\!=\!5$. }
\end{figure}

As per the URLLC users, pilot contamination heavily affects the network availability when $\tp\!<\!16$, especially when SPC is employed and despite a long slot lowers the rate requirements, as we can observe in~\Figref{fig:Fig4_b}.        
Pilot contamination among URLLC users is destructive mainly because they are likely to be close to the BS and to each other, experiencing strong interference that cannot be resolved when their channel estimates are correlated. Hence, our approach aiming at prioritizing the URLLC users in the pilot assignment is technically sound. In addition, increasing the estimation overhead deeply penalizes the network availability since more resources are subtracted to the data transmission, namely the slot length reduces and, as already explained earlier, the rate requirements of the URLLC users increase.  

\begin{figure}
\centerline{\includegraphics[width=\columnwidth]{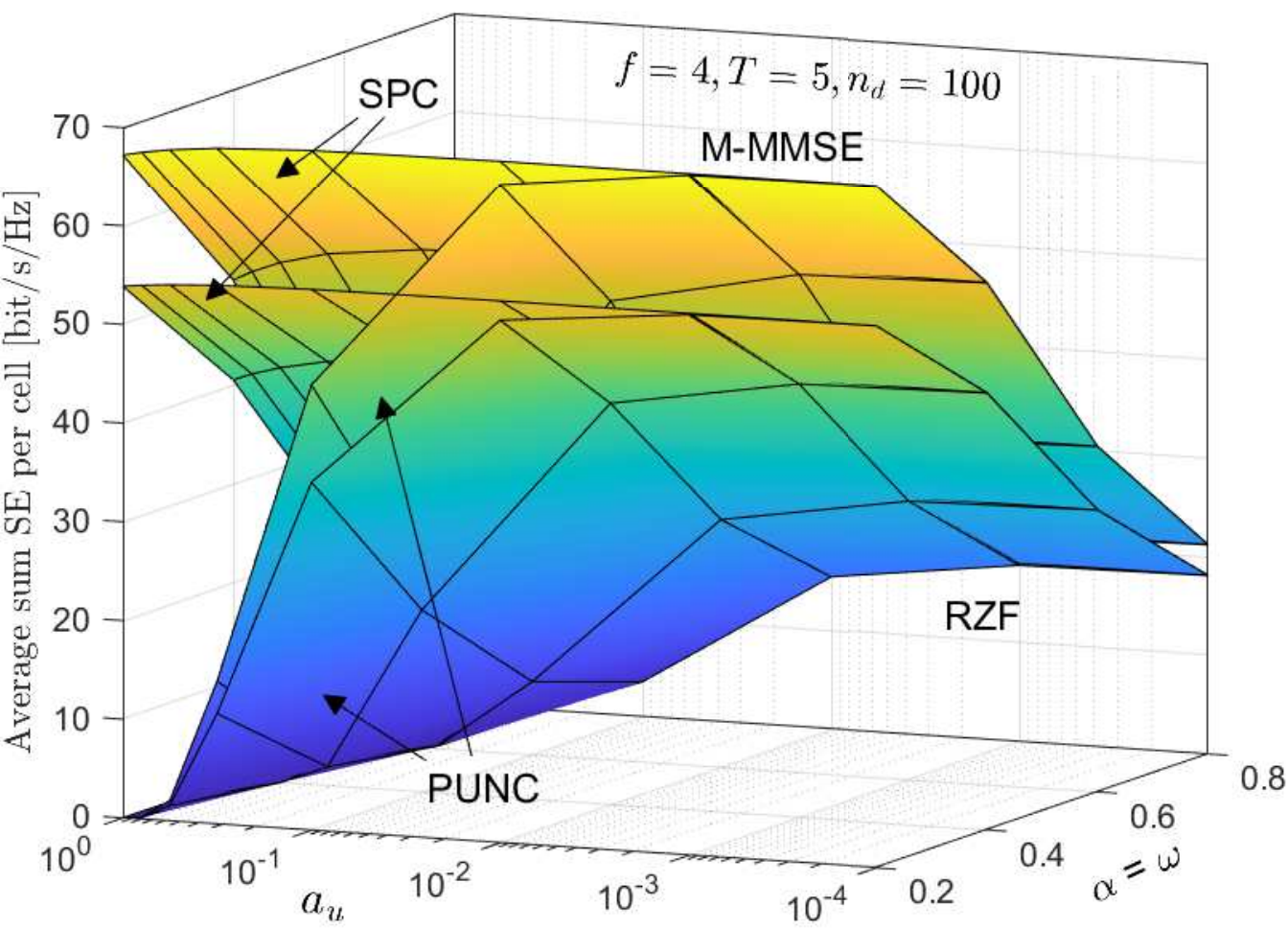}}
\caption{\label{fig:Fig5_a} Average SE per cell, for different transmission and precoding strategies, as $\au$ and $\alpha$ vary. The average is taken over 200 network snapshots. Settings: FPA with $\nu\!=\!0$ and $\omega\!=\!\alpha$, $K\!=\!20$, $\tc\!=\!580$, $f\!=\!4$, $T\!=\!5$, $\nd\!=\!100$.}
\end{figure}

\begin{figure}
\centerline{\includegraphics[width=\columnwidth]{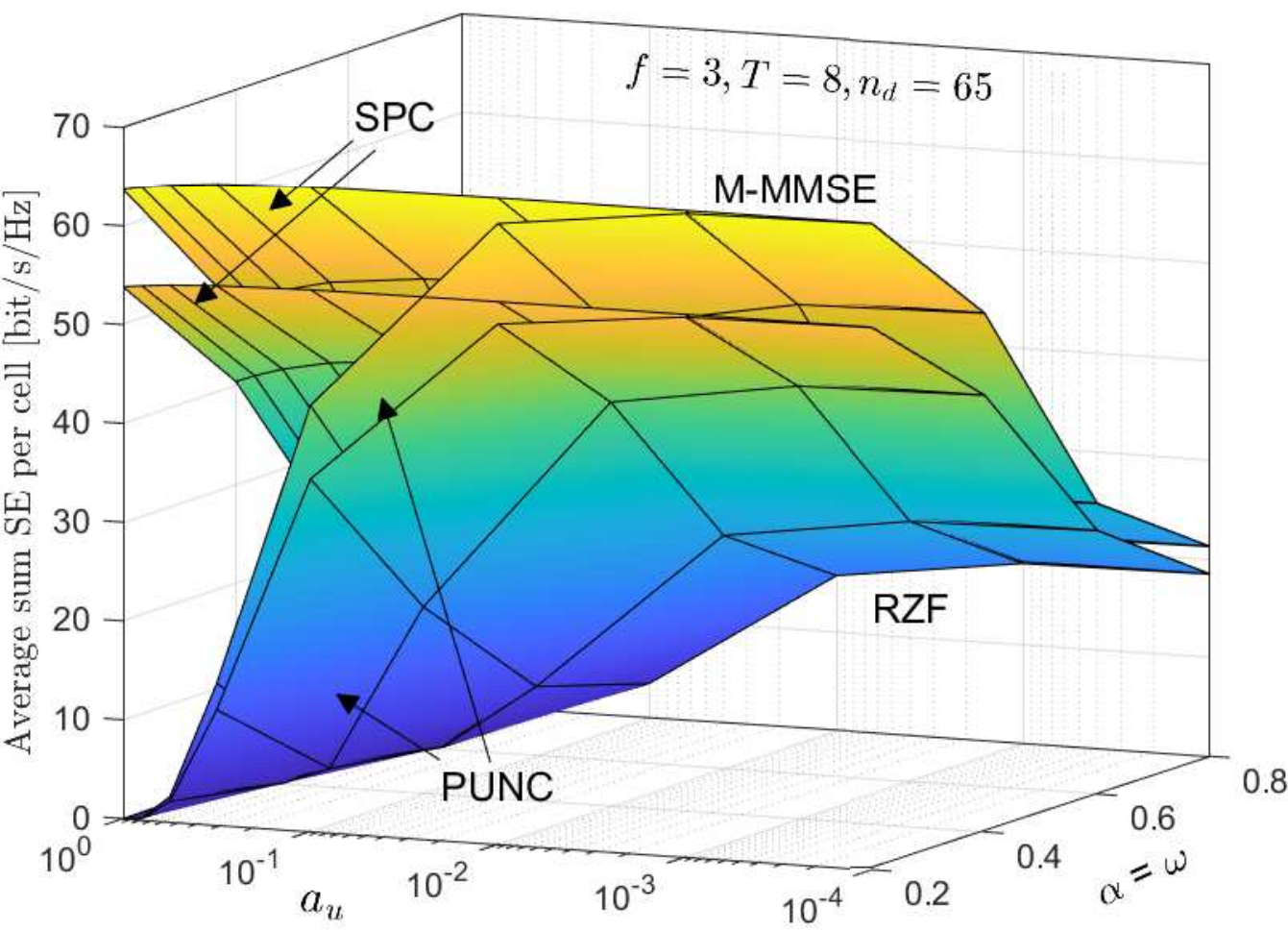}}
\caption{\label{fig:Fig5_a_bis} Average SE per cell, for different transmission and precoding strategies, as $\au$ and $\alpha$ vary. The average is taken over 200 network snapshots. Settings: FPA with $\nu\!=\!0$ and $\omega\!=\!\alpha$, $K\!=\!20$, $\tc\!=\!580$, $f\!=\!3$, $T\!=\!8$, $\nd\!=\!65$.}
\end{figure}

Next we study how the performance are affected by the random activation pattern and the number of potentially active URLLC users per frame. \Figref{fig:Fig5_a} shows the average sum SE per cell as $\au$ and $\alpha$ vary, assuming different transmission and precoding schemes, and FPA with $\nu\!=\!0$ and $\omega\!=\!\alpha$. Notice that, proportionally increasing $\omega$ to $\alpha$ is a reasonable approach for SPC as more power is allocated to an increasing number of potentially active URLLC users, especially for large values of $\au$.
In these simulations, we assume two TDD frame configurations: $(i)$ $f\!=\!4$, $T\!=\!5$, $\nd\!=\!100$, and $(ii)$ $f\!=\!3$, $T\!=\!8$, $\nd\!=\!65$ (whose results are instead shown in~\Figref{fig:Fig5_a_bis}).
First, we observe that similar average sum SE per cell can be achieved by adopting the considered TDD frame configurations: pilot contamination is what slightly degrades the performance of the eMBB users when using the second frame configuration. The performance of PUNC converges to that of SPC when $\au\!\geq\!10^{-2}$, hence for sparse activation patterns, as expected. Again, the performance gap between RZF and M-MMSE reduces in the second scenario (\Figref{fig:Fig5_a_bis}) as the inter-cell pilot contamination decreases the ability of M-MMSE in suppressing the multi-user interference. PUNC provides eMBB service outage for large values of $\au$, whereas SPC is still able to cancel the URLLC user interference and to provide excellent SEs. Lastly, we observe that if the 80\% of the users requests URLLC, then the performance of the eMBB users is reduced of almost one third with respect to the case $\alpha\!=\!0.2$. This result is mainly due to the chosen value of $\omega$ in the FPA scheme that aims to favor the URLLC performance as the number of URLLC users increases. 

\begin{figure}
\centerline{\includegraphics[width=\columnwidth]{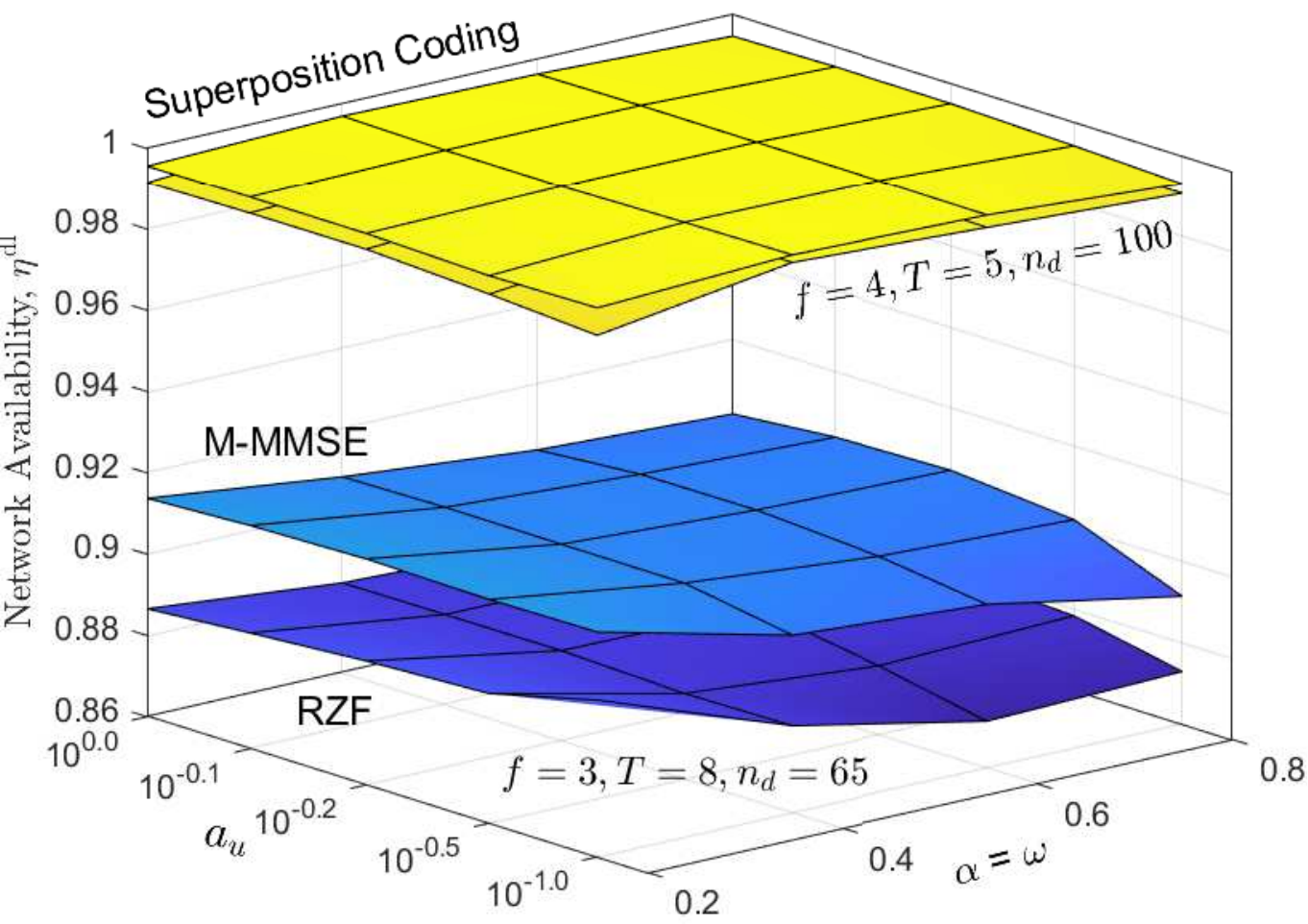}}
\caption{\label{fig:Fig5_c} Network availability, for different precoding strategies, as $\au$ and $\alpha$ vary. The average is taken over 200 network snapshots. Settings: SPC and FPA with $\nu\!=\!0$ and $\omega\!=\!\alpha$, $K\!=\!20$, $\tc\!=\!580$. Two TDD frame configurations are considered.}
\end{figure}

\begin{figure}
\centerline{\includegraphics[width=\columnwidth]{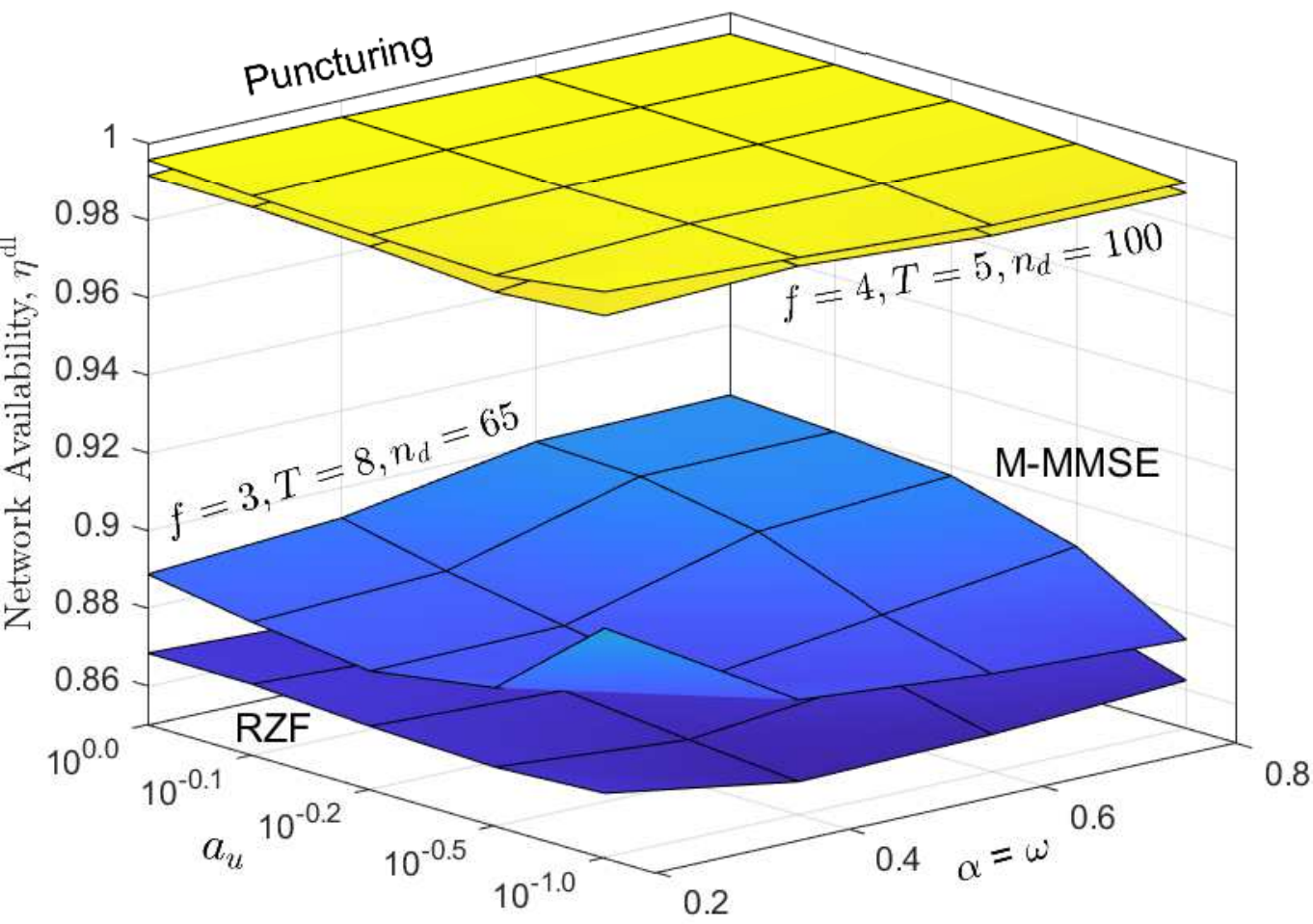}}
\caption{\label{fig:Fig5_d} Network availability, for different precoding strategies, as $\au$ and $\alpha$ vary. Settings: PUNC and FPA with $\nu\!=\!0$ and $\omega\!=\!\alpha$, $K\!=\!20$, $\tc\!=\!580$. Two TDD frame configurations are considered.}
\end{figure}

The performance achieved by the two considered TDD frame configurations appreciably differ in terms of network availability as shown in~\Figref{fig:Fig5_c} for SPC and~\Figref{fig:Fig5_d} for PUNC. 
In both cases, reducing the length of the slot leads to about a 10\% performance loss, while the pilot contamination only concerns the eMBB users. This performance gap is slightly more pronounced when using PUNC because the entire BS power is distributed among the URLLC users causing stronger mutual interference. Overall, the first TDD frame configuration turns to be quite robust to any of the considered transmission and precoding strategies, considered random URLLC activation pattern and URLLC user load.    

A final aspect to be analyzed for this set of simulations is how the probability of eMBB service outage varies with $\au$ and $\alpha$ when PUNC is adopted. This would complete the picture on which operating points PUNC is an effective choice for the eMBB users too, and importantly, further remark the relevance of properly structuring the TDD frame.   
\begin{figure}
\centerline{\includegraphics[width=\columnwidth]{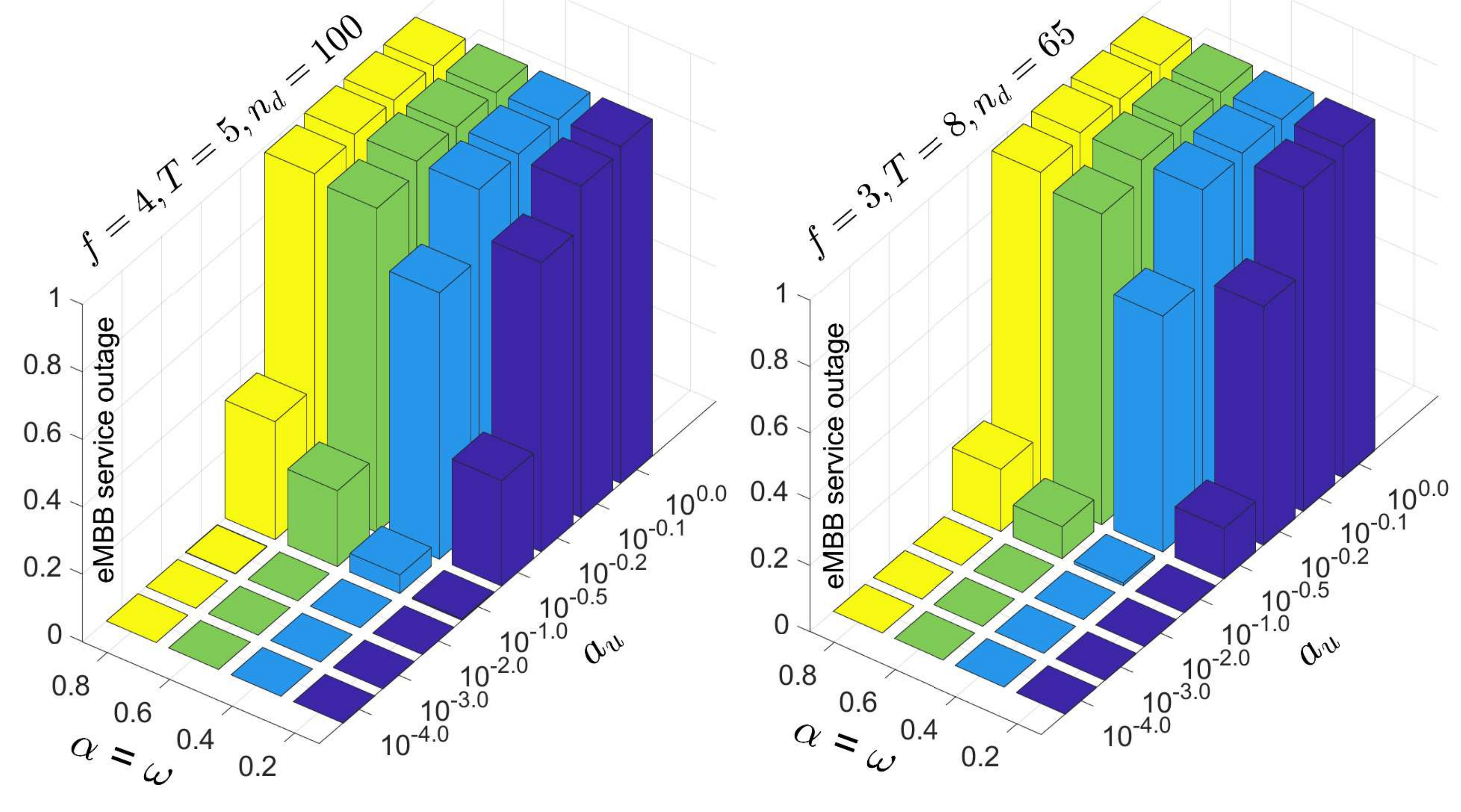}}\caption{\label{fig:Fig5_b} eMBB service outage, for different precoding strategies, as $\au$ and $\alpha$ vary. Settings: PUNC and FPA with $\nu\!=\!0$ and $\omega\!=\!\alpha$, $K\!=\!20$, $\tc\!=\!580$. Two TDD frame configurations are considered.}
\end{figure}
As we can see in~\Figref{fig:Fig5_b}, the advantage of adopting the TDD frame configuration with $T\!=\!8$ slots, when using PUNC, consists in better preventing the eMBB service outage than the configuration with $T\!=\!5$. For instance, when $\au\!=\!10^{-1}$ and $\alpha\!=\!0.8$ or $\alpha\!=\!0.6$, partitioning the share of the frame devoted to the data transmission in 8 slots enables to halve the eMBB outage service compared to the case where 5 slots are adopted. Overall, PUNC can compete with SPC only in scenarios with low URLLC traffic loads, upon properly structuring the TDD frame, as long as a moderate eMBB performance loss is tolerated, either in terms of sum SE per cell or of eMBB service outage. On the other hand, SPC hinges on precoding schemes able to suppress the multi-user interference which, in turn, leverages the spatial degrees of freedom available at the BS and the high accuracy of the acquired CSI.     

\begin{figure}
\centerline{\includegraphics[width=\columnwidth]{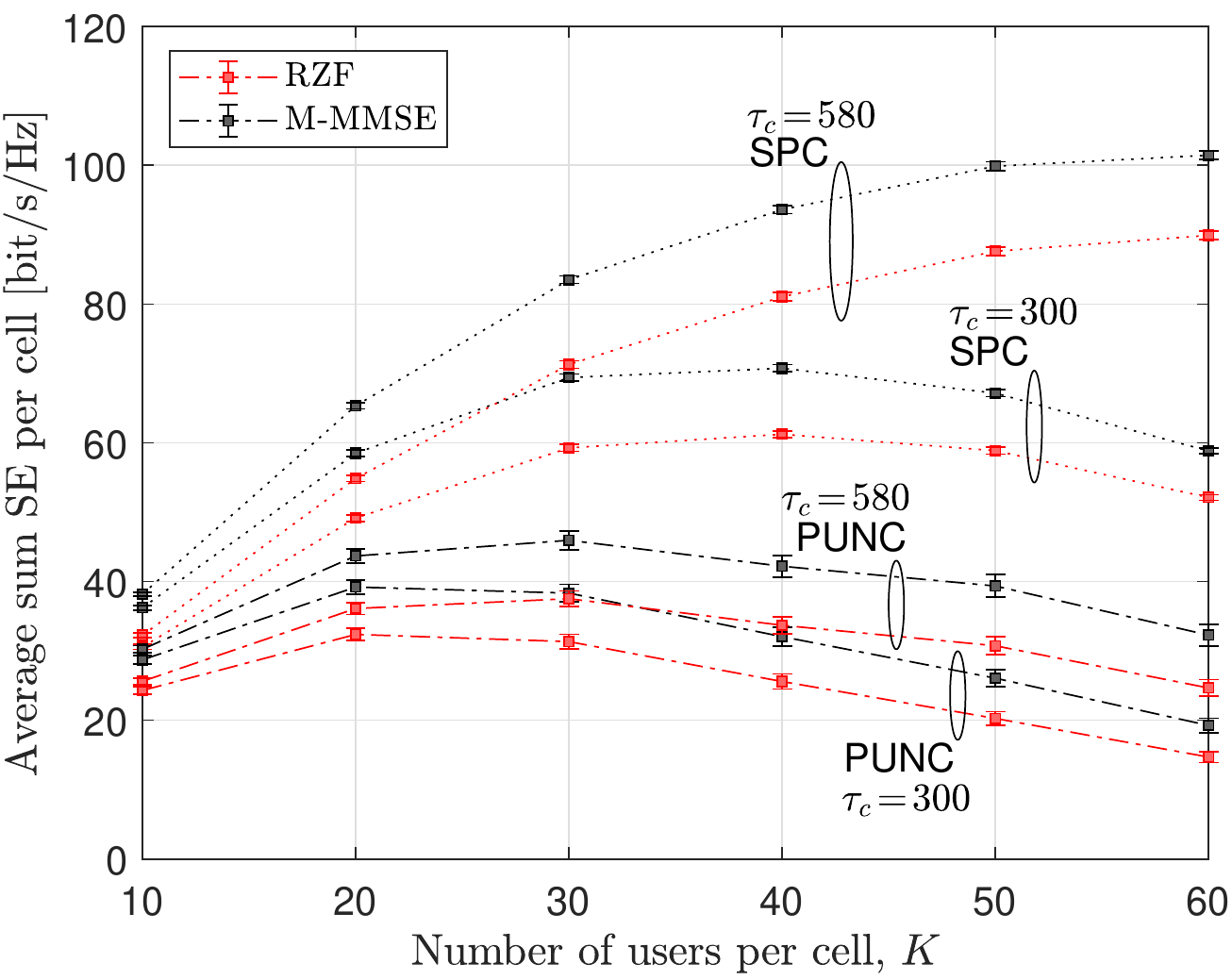} }
\caption{\label{fig:Fig6_a} Average SE per cell (with 95\% confidence interval), for different transmission and precoding strategies, as $K$ and $\tc$ vary. The average is taken over 200 network snapshots. Settings: FPA with $\nu\!=\!0$ and $\omega\!=\!0.2$, $\alpha\!=\!0.2$, $\au\!=\!10^{-1}$, $f\!=\!3$, $T\!=\!5$.  }
\end{figure}

Finally, we evaluate the performance varying the total number of users and the TDD frame length. \Figref{fig:Fig6_a} shows the average sum SE per cell, for different transmission and precoding strategies, as the number of users per cell, $K$, grows from 10 to 60, and considering two different TDD frame lengths, namely 580 and 300 channel uses. The latter may support a shorter coherence time and a narrower coherence bandwidth as well as a higher user mobility compared to the case with 580 channel uses. However, a shorter frame entails less resources that can be allocated to the data transmission and uplink training.        
In these simulations we assume FPA with $\nu\!=\!0$ and $\omega\!=\!0.2$, $\alpha\!=\!0.2$, $\au\!=\!10^{-1}$, $T\!=\!5$ and pilot reuse factor $f\!=\!3$. Moreover, as $\tp\!=\!fK$ and $\tc$ is fixed, for each value of $K$ we have different configurations of uplink training and slot length, i.e., $\tp$ and $\nd$, respectively. From~\Figref{fig:Fig6_a} we observe the average sum SE per cell increasing with $K$, which demonstrates the great ability of SPC with M-MMSE and RZF to spatially multiplex the users. The average sum SE per cell saturates for values of $K$ larger than 60 for $\tc\!=\!580$, and around 40 for $\tc\!=\!300$ wherein the channel estimation overhead heavily burden the SE. PUNC is far inferior to SPC because allocates less resources to the eMBB users and the performance gap increases with $K$ as the number of URLLC users per cell grows proportionally. Therefore, letting $K$ increase makes punctured slots more likely, which not only subtracts resources to the eMBB user reducing its SE but also increases the eMBB service outage, as shown in~\Tableref{tab:res-tp580-fig6}. 
Notice that, the eMBB service outage does not change when varying $\tc$ as long as $T$ is fixed. 

\Tableref{tab:res-tp580-fig6} and~\Tableref{tab:res-tp300-fig6} show the network availability for different transmission and precoding strategies, and different values of $K$, also emphasizing   how $\tp$ and $\nd$ vary accordingly to meet the TDD frame length. In particular, \Tableref{tab:res-tp580-fig6} shows the performance achieved by considering $\tc\!=\!580$, while \Tableref{tab:res-tp300-fig6} shows the performance achieved with $\tc\!=\!300$.  
\begin{table}[!t]
\caption{Network Availability and eMBB service outage, $\tc\!=\!580$}
\centering
\resizebox{\columnwidth}{!}{%
\begin{tabular}{|c|c|c|cccc|c|}
\hline
\multirow{3}{*}{$K$} & \multirow{3}{*}{$\tp$} & \multirow{3}{*}{$\nd$} & \multicolumn{4}{c|}{$\eta^{\mathsf{dl}}$}                                                        & $\varsigma_{\, \mathrm{out}}$ \\ \cline{4-8} 
                     &                        &                        & \multicolumn{2}{c|}{SPC}                                  & \multicolumn{2}{c|}{PUNC}            & \multirow{2}{*}{PUNC}      \\ \cline{4-7}
                     &                        &                        & \multicolumn{1}{c|}{M-MMSE} & \multicolumn{1}{c|}{RZF}    & \multicolumn{1}{c|}{M-MMSE} & RZF    &                            \\ \hline
10                   & 30                     & 110                    & \multicolumn{1}{c|}{0.9989} & \multicolumn{1}{c|}{0.9966} & \multicolumn{1}{c|}{1}      & 0.9989      & 0.0012                     \\ \hline
20                   & 60                     & 104                    & \multicolumn{1}{c|}{0.9988} & \multicolumn{1}{c|}{0.9957} & \multicolumn{1}{c|}{0.9944} & 0.9906 & 0.0038                     \\ \hline
30                   & 90                     & 98                     & \multicolumn{1}{c|}{0.9988} & \multicolumn{1}{c|}{0.9950} & \multicolumn{1}{c|}{0.9934} & 0.9893 & 0.0225                     \\ \hline
40                   & 120                    & 92                     & \multicolumn{1}{c|}{0.9969} & \multicolumn{1}{c|}{0.9885} & \multicolumn{1}{c|}{0.9881} & 0.9819 & 0.0625                     \\ \hline
50                   & 150                    & 86                     & \multicolumn{1}{c|}{0.9864} & \multicolumn{1}{c|}{0.9787} & \multicolumn{1}{c|}{0.9790} & 0.9672 & 0.1050                     \\ \hline
60                   & 180                    & 80                     & \multicolumn{1}{c|}{0.9807} & \multicolumn{1}{c|}{0.9697} & \multicolumn{1}{c|}{0.9728} & 0.9601 & 0.1737                     \\ \hline
\end{tabular}%
}
\label{tab:res-tp580-fig6}
\end{table}
The TDD frame with $\tc\!=\!580$ allows to achieve a network availability above 96\% up to 60 users per cell (of which 12 are URLLC users) with any of the considered transmission and precoding techniques, meaning that such an amount of resources are sufficient to excellently support the considered URLLC user loads and their activation pattern.  
Conversely, the network availability supported by the TDD frame with $\tc\!=\!300$, reported in~\Tableref{tab:res-tp300-fig6}, is considerably lower, even close (or equal) to zero for $K\!\geq\!50$, emphasizing how sensitive the network availability is to the length of the TDD frame, hence to the amount of available resources.
Importantly, we observe the decreasing trend of the network availability as $K$ increases, which for PUNC is milder and mainly due to the shorter URLLC codeword length, but for SPC is severe and mainly due to the increase of the multi-user interference. Indeed, the results in~\Tableref{tab:res-tp300-fig6} clearly confirms that PUNC is more robust than SPC when $K\!\geq\!20$.

\begin{table}[!t]
\caption{Network Availability and eMBB service outage, $\tc\!=\!300$}
\centering
\resizebox{\columnwidth}{!}{%
\begin{tabular}{|c|c|c|cccc|c|}
\hline
\multirow{3}{*}{$K$} & \multirow{3}{*}{$\tp$} & \multirow{3}{*}{$\nd$} & \multicolumn{4}{c|}{$\eta^{\mathsf{dl}}$}                                                        & $\varsigma_{\, \mathrm{out}}$ \\ \cline{4-8} 
                     &                        &                        & \multicolumn{2}{c|}{SPC}                                  & \multicolumn{2}{c|}{PUNC}            & \multirow{2}{*}{PUNC}      \\ \cline{4-7}
                     &                        &                        & \multicolumn{1}{c|}{M-MMSE} & \multicolumn{1}{c|}{RZF}    & \multicolumn{1}{c|}{M-MMSE} & RZF    &                            \\ \hline
10                   & 30                     & 54                     & \multicolumn{1}{c|}{0.7936} & \multicolumn{1}{c|}{0.7683} & \multicolumn{1}{c|}{0.7844} & 0.7534 & 0.0012                     \\ \hline
20                   & 60                     & 48                     & \multicolumn{1}{c|}{0.6786} & \multicolumn{1}{c|}{0.6353} & \multicolumn{1}{c|}{0.6905} & 0.6685 & 0.0038                     \\ \hline
30                   & 90                     & 42                     & \multicolumn{1}{c|}{0.4796} & \multicolumn{1}{c|}{0.4296} & \multicolumn{1}{c|}{0.5646} & 0.5435 & 0.0225                     \\ \hline
40                   & 120                    & 36                     & \multicolumn{1}{c|}{0.1813} & \multicolumn{1}{c|}{0.1457} & \multicolumn{1}{c|}{0.3192} & 0.3192 & 0.0625                     \\ \hline
50                   & 150                    & 30                     & \multicolumn{1}{c|}{0.0021} & \multicolumn{1}{c|}{0}      & \multicolumn{1}{c|}{0.0250} & 0.0250 & 0.1050                     \\ \hline
60                   & 180                    & 24                     & \multicolumn{1}{c|}{0}      & \multicolumn{1}{c|}{0}      & \multicolumn{1}{c|}{0}      & 0      & 0.1737                     \\ \hline
\end{tabular}%
}
\label{tab:res-tp300-fig6}
\end{table}

\section{CONCLUSION}
\label{sec:conclusion}
In this paper, we considered the non-orthogonal multiplexing of heterogeneous services, namely the enhanced mobile broadband (eMBB) and the ultra-reliable low-latency communication (URLLC), in the downlink of a multi-cell massive MIMO system.
eMBB and URLLC have opposite characteristics and diverse requirements. eMBB transmissions involve a large payload that spans multiple radio frames, and demand for high spectral efficiency. While, URLLC users intermittently transmit small payloads in a very short time demanding for low latency and successful probability in the order of $10^{-5}$. 
Such a heterogeneity calls for effective resource allocation strategies to let eMBB and URLLC peacefully coexist. Firstly, we provided a unified information-theoretic framework to assess the spectral efficiency (SE) of the eMBB in the infinite-blocklength ergodic regime, and the error probability of the URLLC in the nonasymptotic finite-blocklength regime. Both analyses encompass imperfect channel state information (CSI) acquisition at the base stations (BSs) via uplink pilot transmissions, pilot contamination and pilot overhead, spatially correlated channels and the lack of CSI at the users. Secondly, we generalized the proposed framework to accommodate two alternative coexistence strategies: puncturing (PUNC) and superposition coding (SPC). The former prevents the inter-service interference aiming to protect the URLLC reliability, while the latter accepts it aiming to maintain the eMBB service.
Thirdly, we numerically evaluated the performance achieved by PUNC and SPC under different precoding and power allocation schemes, and subject to different configurations of the time-division duplex radio frame and URLLC random activation pattern. 
Simulation results revealed that the spatial degrees of freedom available at the BSs, when fully exploited by interference-suppression-based precoding schemes, and upon a high-quality CSI acquisition, enable to significantly resolve the multi-user interference caused by the SPC operation, providing way higher eMBB SE than PUNC, yet ensuring similar great levels of error probability for the URLLC. However, whenever these conditions does not hold, e.g., when a severe pilot contamination degrades the channel estimates or the degrees of freedom are not sufficient to handle the interference between many users, PUNC 
turns to be a necessary operation to preserve the URLLC performance, although it might cause eMBB service outage. Unlike prior works wherein the URLLC performance is inappropriately assessed by using the outage capacity analysis or the error probability obtained by the normal approximation, in this work the finite-blocklength information-theoretic analysis relies on mismatched receivers and on the saddlepoint approximation which is proper of URLLC scenarios in massive MIMO operation. This work can be extended by including massive machine-type communication (mMTC) in the coexistence strategies, and by including the study of the uplink in the proposed generalized framework. Finally, investigating the non-orthogonal multiplexing of heterogeneous services in distributed user-centric systems, such as cell-free massive MIMO~\cite{buzzi2017user,Interdonato2019,Buzzi2019c}, able to provide user's proximity, macrodiversity and ubiquitous connectivity, is certainly an appealing future research direction.

\bibliographystyle{IEEEtran}
\bibliography{IEEEabrv,refs-abbr,refs-slicing,refs-kappa}

\end{document}

%% file: my-macro.tex
\usepackage{mathtools} 
\usepackage{xspace}

\newcommand{\subtext}[1]{\text{\fontfamily{cmr}\fontshape{n}\fontseries{m}\selectfont{}#1}}
\newcommand{\sub}[1]{\ensuremath{_{\subtext{#1}}}} 

\newcommand{\Eqref}[1]{Equation~(\ref{#1})}

\newcommand{\Figref}[1]{Fig.~\ref{#1}}

\newcommand{\Secref}[1]{Section~\ref{#1}}
\newcommand{\Tableref}[1]{Table~\ref{#1}}

\newcommand{\ba}{\mathbf{a}}

\newcommand{\bh}{\mathbf{h}}

\newcommand{\bq}{\mathbf{q}}

\newcommand{\bv}{\mathbf{v}}
\newcommand{\bw}{\mathbf{w}}
\newcommand{\by}{\mathbf{y}}
\newcommand{\bx}{\mathbf{x}}

\newcommand{\bA}{\mathbf{A}}

\newcommand{\bC}{\mathbf{C}}

\newcommand{\bH}{\mathbf{H}}
\newcommand{\bI}{\mathbf{I}}

\newcommand{\bN}{\mathbf{N}}
\newcommand{\bP}{\mathbf{P}}

\newcommand{\bR}{\mathbf{R}}

\newcommand{\bY}{\mathbf{Y}}

\newcommand{\bphi}{\boldsymbol{\phi}}

\newcommand{\bPsi}{\boldsymbol{\Psi}}

\newcommand{\bzero}{\boldsymbol{0}}

\newcommand{\diag}{\text{diag}}

\newcommand{\herm}{^{\mathsf{H}}}
\newcommand{\trans}{^\mathsf{T}}
\newcommand{\inv}{^{-1}}
\newcommand{\conj}{^{*}}

\DeclareMathOperator{\tr}{tr}
\DeclareMathOperator{\E}{\mathsf{E}}

\DeclareMathOperator{\Pro}{\mathsf{Pr}}

\newcommand{\EX}[1]{\E\left\{{#1}\right\}}
\newcommand{\EXs}[2]{\E_{{#1}}\left\{{#2}\right\}}

\newcommand{\Prx}[1]{\Pro\left\{{#1}\right\}}

\DeclareMathOperator*{\argmin}{arg\,min}

\newcommand{\norm}[1]{{ \left\Vert #1 \right\Vert }}

\newcommand{\mC}{\mathbb{C}}
\newcommand{\mR}{\mathbb{R}}

\newcommand{\setP}{\mathcal{P}}

\newcommand{\tc}{\tau_{\mathrm{c}}}
\newcommand{\tp}{\tau_{\mathrm{p}}}
\newcommand{\td}{\tau_{\mathrm{d}}}
\newcommand{\tu}{\tau_{\mathrm{u}}}

\newcommand{\Bc}{B_{\mathrm{c}}}
\newcommand{\Tc}{T_{\mathrm{c}}}

\newcommand{\nd}{n_{\mathrm{d}}}
\newcommand{\nvu}{\sigma^2_{\mathrm{u}}}
\newcommand{\nvd}{\sigma^2_{\mathrm{d}}}
\newcommand{\sigmash}{\sigma_{\mathsf{sh}}}
\newcommand{\pilot}{^{\mathsf{p}}}

\newcommand{\hhat}{\widehat{\bh}}
\newcommand{\htilde}{\widetilde{\bh}}
\newcommand{\iid}{\text{i.i.d.}}
\newcommand{\CG}[2]{\mathcal{CN}\left({#1},{#2}\right)}
\newcommand{\G}[2]{\mathcal{N}\left({#1},{#2}\right)}

\def\Qexp{\Psi_{\nd,\varepsilon}}
\newcommand{\Atj}{\tilde{A}^t_j}
\newcommand{\Atl}{\tilde{A}^t_l}
\newcommand{\Atji}{{A}^t_{ji}}
\newcommand{\Atjk}{{A}^t_{jk}}
\newcommand{\Atju}{{A}^t_{ju}}
\newcommand{\rhoe}{\rho^{\mathsf{e}}}
\newcommand{\rhou}{\rho^{\mathsf{u}}}
\newcommand{\rhos}{\rho^{\mathsf{s}}}
\newcommand{\rhoMax}{\rho^{\mathsf{max}}}

\newcommand{\Ku}{K_{\mathrm{u}}}
\newcommand{\Ke}{K_{\mathrm{e}}}
\newcommand{\setKe}{\mathcal{K}^{\mathrm{e}}}
\newcommand{\setKu}{\mathcal{K}^{\mathrm{u}}}
\newcommand{\au}{a_{\mathrm{u}}}
\newcommand{\datae}{\varsigma^{\mathsf{e}}}
\newcommand{\datau}{\varsigma^{\mathsf{u}}}
\newcommand{\datas}{\varsigma^{\mathsf{s}}}

\makeatletter
\def\@setsize#1#2#3#4{
    \@nomath#1
    \let\@currsize#1
    \baselineskip #2
    \baselineskip \baselinestretch\baselineskip
    \parskip \baselinestretch\parskip
    \setbox\strutbox \hbox{
        \vrule height.7\baselineskip
            depth.3\baselineskip
            width\z@}
    \skip\footins \baselinestretch\skip\footins
    \normalbaselineskip\baselineskip#3#4}
\makeatother

\makeatletter
\newcommand{\setstretch}[1]{
    \def\baselinestretch{#1}%
    \@currsize
    }
\makeatother

%% file: TDDframe.tex
\usetikzlibrary{patterns}

 
\tikzset{
pattern size/.store in=\mcSize, 
pattern size = 5pt,
pattern thickness/.store in=\mcThickness, 
pattern thickness = 0.3pt,
pattern radius/.store in=\mcRadius, 
pattern radius = 1pt}
\makeatletter
\pgfutil@ifundefined{pgf@pattern@name@_fdcpbizmp}{
\pgfdeclarepatternformonly[\mcThickness,\mcSize]{_fdcpbizmp}
{\pgfqpoint{0pt}{-\mcThickness}}
{\pgfpoint{\mcSize}{\mcSize}}
{\pgfpoint{\mcSize}{\mcSize}}
{
\pgfsetcolor{\tikz@pattern@color}
\pgfsetlinewidth{\mcThickness}
\pgfpathmoveto{\pgfqpoint{0pt}{\mcSize}}
\pgfpathlineto{\pgfpoint{\mcSize+\mcThickness}{-\mcThickness}}
\pgfusepath{stroke}
}}
\makeatother
\tikzset{every picture/.style={line width=0.75pt}} 

\begin{tikzpicture}[x=0.75pt,y=0.75pt,yscale=-.9,xscale=.9]

\draw  [draw opacity=0][fill={rgb, 255:red, 211; green, 245; blue, 243 }  ,fill opacity=1 ] (171.5,102.5) -- (380,102.5) -- (380,140) -- (171.5,140) -- cycle ;
\draw  [draw opacity=0][fill={rgb, 255:red, 241; green, 236; blue, 170 }  ,fill opacity=1 ] (101.5,102) -- (171,102) -- (171,139.5) -- (101.5,139.5) -- cycle ;
\draw  [draw opacity=0] (100.5,101) -- (381,101) -- (381,141.5) -- (100.5,141.5) -- cycle ; \draw  [color={rgb, 255:red, 214; green, 208; blue, 208 }  ,draw opacity=1 ] (100.5,101) -- (100.5,141.5)(110.5,101) -- (110.5,141.5)(120.5,101) -- (120.5,141.5)(130.5,101) -- (130.5,141.5)(140.5,101) -- (140.5,141.5)(150.5,101) -- (150.5,141.5)(160.5,101) -- (160.5,141.5)(170.5,101) -- (170.5,141.5)(180.5,101) -- (180.5,141.5)(190.5,101) -- (190.5,141.5)(200.5,101) -- (200.5,141.5)(210.5,101) -- (210.5,141.5)(220.5,101) -- (220.5,141.5)(230.5,101) -- (230.5,141.5)(240.5,101) -- (240.5,141.5)(250.5,101) -- (250.5,141.5)(260.5,101) -- (260.5,141.5)(270.5,101) -- (270.5,141.5)(280.5,101) -- (280.5,141.5)(290.5,101) -- (290.5,141.5)(300.5,101) -- (300.5,141.5)(310.5,101) -- (310.5,141.5)(320.5,101) -- (320.5,141.5)(330.5,101) -- (330.5,141.5)(340.5,101) -- (340.5,141.5)(350.5,101) -- (350.5,141.5)(360.5,101) -- (360.5,141.5)(370.5,101) -- (370.5,141.5)(380.5,101) -- (380.5,141.5) ; \draw  [color={rgb, 255:red, 214; green, 208; blue, 208 }  ,draw opacity=1 ] (100.5,101) -- (381,101)(100.5,111) -- (381,111)(100.5,121) -- (381,121)(100.5,131) -- (381,131)(100.5,141) -- (381,141) ; \draw  [color={rgb, 255:red, 214; green, 208; blue, 208 }  ,draw opacity=1 ]  ;
\draw  [draw opacity=0][fill={rgb, 255:red, 184; green, 233; blue, 134 }  ,fill opacity=0.5 ] (157.5,189.5) -- (187.5,189.5) -- (187.5,229.5) -- (157.5,229.5) -- cycle ;
\draw  [draw opacity=0][fill={rgb, 255:red, 216; green, 107; blue, 119 }  ,fill opacity=0.5 ] (365.5,188) -- (395.5,188) -- (395.5,228) -- (365.5,228) -- cycle ;
\draw   (100.5,101) -- (380.5,101) -- (380.5,141) -- (100.5,141) -- cycle ;
\draw    (351,101) -- (351,141) ;
\draw    (201,100.5) -- (201,140.5) ;
\draw    (231,101) -- (231,141) ;
\draw    (260.5,101) -- (260.5,141) ;
\draw    (290.5,101) -- (290.5,141) ;
\draw    (320.5,101) -- (320.5,141) ;
\draw    (119,158) .. controls (123.41,156.04) and (112.93,165.12) .. (126.63,139.14) ;
\draw [shift={(127.5,137.5)}, rotate = 118.18] [color={rgb, 255:red, 0; green, 0; blue, 0 }  ][line width=0.75]    (10.93,-3.29) .. controls (6.95,-1.4) and (3.31,-0.3) .. (0,0) .. controls (3.31,0.3) and (6.95,1.4) .. (10.93,3.29)   ;
\draw   (384,141) .. controls (388.67,141) and (391,138.67) .. (391,134) -- (391,130) .. controls (391,123.33) and (393.33,120) .. (398,120) .. controls (393.33,120) and (391,116.67) .. (391,110)(391,113) -- (391,108) .. controls (391,103.33) and (388.67,101) .. (384,101) ;
\draw    (103.5,77.5) -- (377,77.5) ;
\draw [shift={(380,77.5)}, rotate = 180] [fill={rgb, 255:red, 0; green, 0; blue, 0 }  ][line width=0.08]  [draw opacity=0] (8.93,-4.29) -- (0,0) -- (8.93,4.29) -- cycle    ;
\draw [shift={(100.5,77.5)}, rotate = 0] [fill={rgb, 255:red, 0; green, 0; blue, 0 }  ][line width=0.08]  [draw opacity=0] (8.93,-4.29) -- (0,0) -- (8.93,4.29) -- cycle    ;
\draw    (103.5,94.5) -- (168.5,94.5) ;
\draw [shift={(171.5,94.5)}, rotate = 180] [fill={rgb, 255:red, 0; green, 0; blue, 0 }  ][line width=0.08]  [draw opacity=0] (8.93,-4.29) -- (0,0) -- (8.93,4.29) -- cycle    ;
\draw [shift={(100.5,94.5)}, rotate = 0] [fill={rgb, 255:red, 0; green, 0; blue, 0 }  ][line width=0.08]  [draw opacity=0] (8.93,-4.29) -- (0,0) -- (8.93,4.29) -- cycle    ;
\draw    (174.5,94.5) -- (377,94.5) ;
\draw [shift={(380,94.5)}, rotate = 180] [fill={rgb, 255:red, 0; green, 0; blue, 0 }  ][line width=0.08]  [draw opacity=0] (8.93,-4.29) -- (0,0) -- (8.93,4.29) -- cycle    ;
\draw [shift={(171.5,94.5)}, rotate = 0] [fill={rgb, 255:red, 0; green, 0; blue, 0 }  ][line width=0.08]  [draw opacity=0] (8.93,-4.29) -- (0,0) -- (8.93,4.29) -- cycle    ;
\draw  [draw opacity=0][pattern=_fdcpbizmp,pattern size=3.9000000000000004pt,pattern thickness=0.75pt,pattern radius=0pt, pattern color={rgb, 255:red, 0; green, 0; blue, 0}] (352.5,102.5) -- (379.25,102.5) -- (379.25,140) -- (352.5,140) -- cycle ;
\draw    (377,158) .. controls (364.81,149.71) and (365.46,147.13) .. (365.03,122.44) ;
\draw [shift={(365,120.5)}, rotate = 88.92] [color={rgb, 255:red, 0; green, 0; blue, 0 }  ][line width=0.75]    (10.93,-3.29) .. controls (6.95,-1.4) and (3.31,-0.3) .. (0,0) .. controls (3.31,0.3) and (6.95,1.4) .. (10.93,3.29)   ;
\draw  [dash pattern={on 0.84pt off 2.51pt}]  (304.5,120.5) -- (329,180) ;
\draw  [dash pattern={on 0.84pt off 2.51pt}]  (245.5,120) -- (225,179.5) ;
\draw  [draw opacity=0][fill={rgb, 255:red, 216; green, 107; blue, 119 }  ,fill opacity=0.5 ] (205,189.5) -- (235,189.5) -- (235,229.5) -- (205,229.5) -- cycle ;
\draw  [draw opacity=0] (205,189.5) -- (235.5,189.5) -- (235.5,230) -- (205,230) -- cycle ; \draw   (205,189.5) -- (205,230)(215,189.5) -- (215,230)(225,189.5) -- (225,230)(235,189.5) -- (235,230) ; \draw   (205,189.5) -- (235.5,189.5)(205,199.5) -- (235.5,199.5)(205,209.5) -- (235.5,209.5)(205,219.5) -- (235.5,219.5)(205,229.5) -- (235.5,229.5) ; \draw    ;
\draw  [draw opacity=0] (157.5,189.5) -- (188,189.5) -- (188,230) -- (157.5,230) -- cycle ; \draw   (157.5,189.5) -- (157.5,230)(167.5,189.5) -- (167.5,230)(177.5,189.5) -- (177.5,230)(187.5,189.5) -- (187.5,230) ; \draw   (157.5,189.5) -- (188,189.5)(157.5,199.5) -- (188,199.5)(157.5,209.5) -- (188,209.5)(157.5,219.5) -- (188,219.5)(157.5,229.5) -- (188,229.5) ; \draw    ;
\draw  [draw opacity=0][fill={rgb, 255:red, 184; green, 233; blue, 134 }  ,fill opacity=0.5 ] (365.5,188) -- (395.5,188) -- (395.5,228) -- (365.5,228) -- cycle ;
\draw  [draw opacity=0][fill={rgb, 255:red, 184; green, 233; blue, 134 }  ,fill opacity=0.5 ] (276,188.5) -- (306,188.5) -- (306,228.5) -- (276,228.5) -- cycle ;
\draw  [draw opacity=0][fill={rgb, 255:red, 216; green, 107; blue, 119 }  ,fill opacity=0.5 ] (320.5,188.5) -- (350.5,188.5) -- (350.5,228.5) -- (320.5,228.5) -- cycle ;
\draw  [draw opacity=0] (320.5,188.5) -- (351,188.5) -- (351,229) -- (320.5,229) -- cycle ; \draw   (320.5,188.5) -- (320.5,229)(330.5,188.5) -- (330.5,229)(340.5,188.5) -- (340.5,229)(350.5,188.5) -- (350.5,229) ; \draw   (320.5,188.5) -- (351,188.5)(320.5,198.5) -- (351,198.5)(320.5,208.5) -- (351,208.5)(320.5,218.5) -- (351,218.5)(320.5,228.5) -- (351,228.5) ; \draw    ;
\draw  [draw opacity=0] (276,188.5) -- (306.5,188.5) -- (306.5,229) -- (276,229) -- cycle ; \draw   (276,188.5) -- (276,229)(286,188.5) -- (286,229)(296,188.5) -- (296,229)(306,188.5) -- (306,229) ; \draw   (276,188.5) -- (306.5,188.5)(276,198.5) -- (306.5,198.5)(276,208.5) -- (306.5,208.5)(276,218.5) -- (306.5,218.5)(276,228.5) -- (306.5,228.5) ; \draw    ;
\draw  [draw opacity=0] (365.5,188) -- (396,188) -- (396,228.5) -- (365.5,228.5) -- cycle ; \draw   (365.5,188) -- (365.5,228.5)(375.5,188) -- (375.5,228.5)(385.5,188) -- (385.5,228.5)(395.5,188) -- (395.5,228.5) ; \draw   (365.5,188) -- (396,188)(365.5,198) -- (396,198)(365.5,208) -- (396,208)(365.5,218) -- (396,218)(365.5,228) -- (396,228) ; \draw    ;
\draw  [dash pattern={on 0.84pt off 2.51pt}] (141,183.5) -- (253.5,183.5) -- (253.5,249.5) -- (141,249.5) -- cycle ;
\draw  [dash pattern={on 0.84pt off 2.51pt}] (259.5,183) -- (406.5,183) -- (406.5,250) -- (259.5,250) -- cycle ;

\draw (313,231) node [anchor=north west][inner sep=0.75pt]  [font=\footnotesize] [align=left] {URLLC};
\draw (271.5,231) node [anchor=north west][inner sep=0.75pt]  [font=\footnotesize] [align=left] {eMBB};
\draw (198.5,231) node [anchor=north west][inner sep=0.75pt]  [font=\footnotesize] [align=left] {URLLC};
\draw (152.5,231) node [anchor=north west][inner sep=0.75pt]  [font=\footnotesize] [align=left] {eMBB};
\draw (178.5,252) node [anchor=north west][inner sep=0.75pt]  [font=\footnotesize] [align=left] {PUNC};
\draw (189,204) node [anchor=north west][inner sep=0.75pt]  [font=\footnotesize] [align=left] {or};
\draw (352,204) node [anchor=north west][inner sep=0.75pt]  [font=\footnotesize] [align=left] {=};
\draw (308,204) node [anchor=north west][inner sep=0.75pt]  [font=\footnotesize] [align=left] {+};
\draw (315.5,252) node [anchor=north west][inner sep=0.75pt]  [font=\footnotesize] [align=left] {SPC};
\draw (377.5,148.5) node [anchor=north west][inner sep=0.75pt]  [font=\footnotesize] [align=left] {one slot, $\displaystyle n_{\mathrm{d}}$};
\draw  [draw opacity=0]  (264,107.5) -- (286,107.5) -- (286,129.5) -- (264,129.5) -- cycle  ;
\draw (265,115) node [anchor=north west][inner sep=0.75pt]  [font=\footnotesize] [align=left] {\begin{minipage}[lt]{12.3pt}\setlength\topsep{0pt}
\begin{center}
$\displaystyle \tau _{\mathrm{d}}$
\end{center}

\end{minipage}};
\draw (220.5,80) node [anchor=north west][inner sep=0.75pt]  [font=\scriptsize] [align=left] {\begin{minipage}[lt]{72.7pt}\setlength\topsep{0pt}
\begin{center}
DL Data Transmission
\end{center}

\end{minipage}};
\draw (107.5,80) node [anchor=north west][inner sep=0.75pt]  [font=\scriptsize] [align=left] {\begin{minipage}[lt]{36.99pt}\setlength\topsep{0pt}
\begin{center}
UL training
\end{center}

\end{minipage}};
\draw  [draw opacity=0]  (126.5,107.5) -- (148.5,107.5) -- (148.5,129.5) -- (126.5,129.5) -- cycle  ;
\draw (127.5,115) node [anchor=north west][inner sep=0.75pt]  [font=\footnotesize] [align=left] {\begin{minipage}[lt]{12.24pt}\setlength\topsep{0pt}
\begin{center}
$\displaystyle \tau _{\mathrm{p}}$
\end{center}

\end{minipage}};
\draw (397.5,93.5) node [anchor=north west][inner sep=0.75pt]  [font=\footnotesize] [align=left] {\begin{minipage}[lt]{44.91pt}\setlength\topsep{0pt}
\begin{center}
Coherence \\bandwidth \\$\displaystyle B_{\mathrm{c}}$
\end{center}

\end{minipage}};
\draw (123,149) node [anchor=north west][inner sep=0.75pt]  [font=\footnotesize] [align=left] {one channel use};
\draw (189,57.5) node [anchor=north west][inner sep=0.75pt]  [font=\footnotesize] [align=left] {Coherence time, $\displaystyle T_{\mathrm{c}}$};

\end{tikzpicture}